\documentclass[fleqn,usenatbib]{mnras}
\usepackage{newtxtext,newtxmath}
\usepackage[T1]{fontenc}

\DeclareRobustCommand{\VAN}[3]{#2}
\let\VANthebibliography\thebibliography
\def\thebibliography{\DeclareRobustCommand{\VAN}[3]{##3}\VANthebibliography}

\newcommand{\pa}{\partial}

\usepackage{graphicx}	% Including figure files
\usepackage{float}
\usepackage{amsmath}	% Advanced maths commands

\title[Warped Disks with Outflows]{Semi-Analytic Solutions For Warped, Super-Eddington Accretion Disks with Outflows}

\author[Gabriel Wohlfarth et al.]{
Gabriel Wohlfarth,$^{1,2}$\thanks{E-mail: gabe.wohlfarth@student.uva.nl}
Matthew Middleton,$^{2}$
Omer Blaes$^{3}$
and P. Chris Fragile$^{1}$
\\
$^{1}$Department of Physics \& Astronomy, College of Charleston, 66 George Street, Charleston, SC, 29424, USA\\
$^{2}$School of Physics \& Astronomy, University of Southampton, Southampton, Southampton SO17 1BJ, UK\\
$^{3}$Physics Department, University of California at Santa Barbara, Santa Barbara, CA, 93106, USA
}

\date{Accepted XXX. Received YYY; in original form ZZZ}

\pubyear{\the\year{}}

\begin{document}
\label{firstpage}
\pagerange{\pageref{firstpage}--\pageref{lastpage}}
\maketitle

\begin{abstract}
The discovery of billion solar mass black holes at high redshift challenges any standard Eddington-limited growth model. Previous work has shown that an advective accretion disk can account for these early black holes with episodic accretion. It has also been shown that if the accretion disk of a black hole is misaligned from the spin axis of the black hole, the disk can become advective within a certain radius. While warped and misaligned accretion disks have been explored in numerical simulations, there is no fully or semi-analytic framework that self-consistently connects disk warp, mass inflow, winds, and energy transport in the super-Eddington regime. In this work, we develop a first-principles semi-analytic model of a radiation-pressure-dominated, warped accretion disk, starting from the coupled conservation equations for mass, energy, and angular momentum. We find profiles for the radial velocity and surface density that explicitly account for disk warping and mass loss and determine the radius at which the disk becomes advective. We explore changes with black hole spin, misalignment angle and viscosity, reproducing relevant results from numerical studies.

\end{abstract}


%%%%%%%%%%%%%%%%%%%%%%%%%%%%%%%%%%%%%%%%%%%%%%%%%%

%%%%%%%%%%%%%%%%% BODY OF PAPER %%%%%%%%%%%%%%%%%%

\section{Introduction}
\label{sec:1}

Observations of quasars \citep[e.g.,][]{Willott_03, Wu_2016} appear to imply that billion solar mass black holes are present less than a billion years after the Big Bang \citep[but see][for issues around mass inference]{King_25}. Taking the masses at face value presents a challenge, as even sustained accretion at or near the Eddington limit is insufficient to grow such objects from stellar remnant (PopIII) black holes within the allowed timeframe. Whilst large amounts of radial advection \citep[e.g.,][]{Jiang_2014}, could potentially solve this issue \citep[and is regularly used within cosmological simulations, e.g.,][]{sijacki}, recent numerical simulations with radially extended super-Eddington discs which start from a Novikov-Thorne solution, have shown that radiative outflows limit long term average accretion rates to around the Eddington limit onto the black hole \citep{Fragile_25}.  

%In contrast, radiation dominated slim-disk models are able to induce advection when the photon diffusion time becomes larger than the mass infall time, allowing the photons to be advected inwards with the gasses. In this regime, accretion can exceed the Eddington limit because the luminosity remains low due to the inward advection of photons. This, in turn, allows the inward gravitational force to dominate, increasing the amount of mass flowing into the black hole. This has been shown to hold true in numerical simulations such as those by \citet{Volonteri_2005}.

One way to remedy the situation is to introduce misalignment between the disk and black hole spin axes, as the resulting warped structure (driven by Lense-Thirring torques) leads to the formation of standing shocks {\citep{Fragile08,Kaaz2023}}
and highly efficient angular momentum transport. This is now starting to be explored in three dimensional, general relativistic radiative magnetohydrodynamic (3D GR-RMHD) simulations \citep{Fragile26} and confirms that higher than Eddington growth is indeed possible. Whilst it is not possible to capture all of the effects of such flows analytically, one can make approximations, compare to the outputs of simulations and explore a large parameter space which is not feasible numerically (at least not presently). 

%In the past, analytical and numerical work has been done showing that inducing a warp within the black hole accretion disk can create an advection dominated region within the disk, allowing for a lower effective luminosity. Also in the past, analytical and numerical work has been done showing that at super-Eddington accretion rates, the disk can be globally super-Eddington, while remaining sub-Eddington locally \citep{Lipunova_99}.

In this work, we extend the analytical framework of super-Eddington accretion with radial advection and mass loss developed by \citet{Lipunova_99} and \citet{Poutanen_07} to include a Lense-Thirring driven disk warp \citep[which is reasonably matched by recent simulations, i.e.,][]{Fragile26}. We then compute viscous heating, radiative cooling, and advective transport within the disk to explore the resulting structure and energy balance, and determine the radius at which advection begins to dominate.

Section \ref{sec:2} presents the full derivation of our semi-analytic system of equations. Section \ref{sec:Results} then presents our results for a range of disk misalignment angles. We wrap up in Section \ref{sec:discussion} with some discussion and conclusions.

\section{Methods}
\label{sec:2}
\subsection{Physical Framework and Assumptions}
\label{sec:2.1}
We construct a semi-analytic, steady-state model for a warped accretion disk around a black hole. We set all time derivatives to zero, allowing the disk structure to be determined analytically rather than through a time-dependent numerical simulation. This assumption significantly reduces computational expense while still allowing us to determine the radial structure of the system. We assume radiation pressure dominates, local thermal equilibrium is maintained, as is a Keplerian angular velocity in the disc. As described by \citet{Pringle_92}, misalignment results in two viscosities: $\nu_1$ which corresponds to the azimuthal shear, and $\nu_2$ which corresponds to the vertical shear. Both of these are expected to be functions of radius. Within the spherization radius, mass loss via winds is explicitly included in the mass and angular momentum conservation equations.

\begin{figure}
    \centering
    \includegraphics[width=1.0\linewidth]{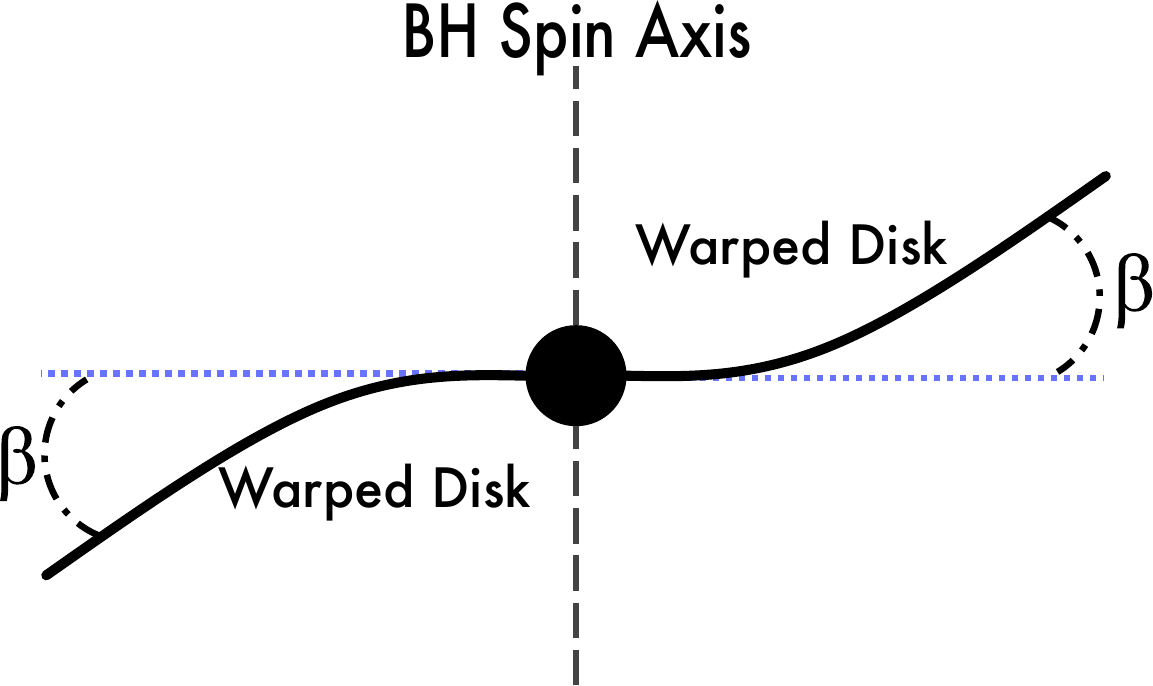}
    \caption{Simple diagram showing the structure of the warped accretion disk. The inner disk becomes aligned due to Lense-Thirring torques while the outer disk remains warped.}
    \label{fig:warp_disk_model}
\end{figure}

\subsection{Conservation Equations for a Warped Disk}
We start from the conservation equations for mass and angular momentum from \citet{Pringle_92}:
\begin{equation}
\label{eq:angmom_con}
\begin{split}
    \frac{1}{R} \frac{\pa}{\partial R} \left(\Sigma V_R R^3 \Omega \vec{l}\right) + \boldsymbol{\frac{R^2 \Omega \vec{l}}{R} \frac{\partial \dot{m}_{\rm w}}{\partial R}} =
    \\ \frac{1}{R} \frac{\partial}{\partial R}\left(\nu_1 \Sigma R^3  \vec{l} \ \frac{\partial \Omega}{\partial R}\right) + \frac{1}{R} \frac{\partial}{\partial R} \left(\frac{1}{2} \nu_2 \Sigma R^3 \Omega \frac{\partial \vec{l}} {\partial R}\right)
\end{split}
\end{equation}
and
\begin{equation}
\label{eq:mass_con}
    \frac{1}{R} \frac{\partial}{\partial R}\left(R \Sigma V_R\right) + \boldsymbol{\frac{1}{R} \frac{\partial \dot{m}_{\rm w}}{\partial R}} = 0
\end{equation}
where $\vec{l}(R)$ is the local unit tilt vector, $\Omega_K(R)$ is Keplerian angular velocity, $V_R(R)$ is the radial velocity, and $\Sigma(R)$ is the surface density of the disk. We have added terms for the mass and angular momentum losses due to outflows to these equations (indicated in bold).

We assume disc warp profiles based on a modified version of the constant viscosity, linear warp solution of \citet{SF96}:
\begin{equation}
    \label{lx}
    l_x = l_{x\infty} \cos \left(\frac{2R_{\rm warp}}{R} \right)\exp \left( {-\frac{2R_{\rm warp}}{R}}\right) ~,
\end{equation}
\begin{equation}
    \label{ly}
    l_y = l_{x\infty} \sin \left( \frac{2R_{\rm warp}}{R} \right)\exp \left({-\frac{2R_{\rm warp}}{R}}\right) ~,
\end{equation}
and
\begin{equation}
    \label{lz}
    l_z = \sqrt{1 - l_x^2 - l_y^2} ~,
\end{equation}
where $l_{x\infty} = \sin \beta$ is the x-component of the tilt vector at a distance of infinity (far enough away so that $\beta$ stops changing) and $R_{\rm warp}$ is given by the expression \citep{NKDJ12}:
\begin{equation}
    R_{\rm warp} = \left( \frac{4}{3} \frac{a |\sin \beta | R}{\alpha \left< H_{\rm flat} \right>} \right)^{2/3}
    \label{Rwarp}
\end{equation}
where $\left< H_{\rm flat} \right>$ is the spatially averaged height of the flat disk, $a$ is the Kerr spin parameter of the black hole, $\beta$ is the angle between the warped disk and the plane perpendicular to the spin axis of the black hole (as shown in Figure \ref{fig:warp_disk_model}), and $\alpha$ is the viscosity parameter. It should be noted that we are adopting the solutions given in equations \ref{lx}-\ref{lz} for the warp profile as opposed to fully solving the angular momentum equation for $\vec{l}(R)$ simultaneously with the disk structure. Equations \ref{eq:angmom_con} and \ref{eq:mass_con} are therefore used to find the profiles for radial transport and mass loss for a given warp profile.

\begin{figure}
    \centering
    \includegraphics[width=1.0\linewidth]{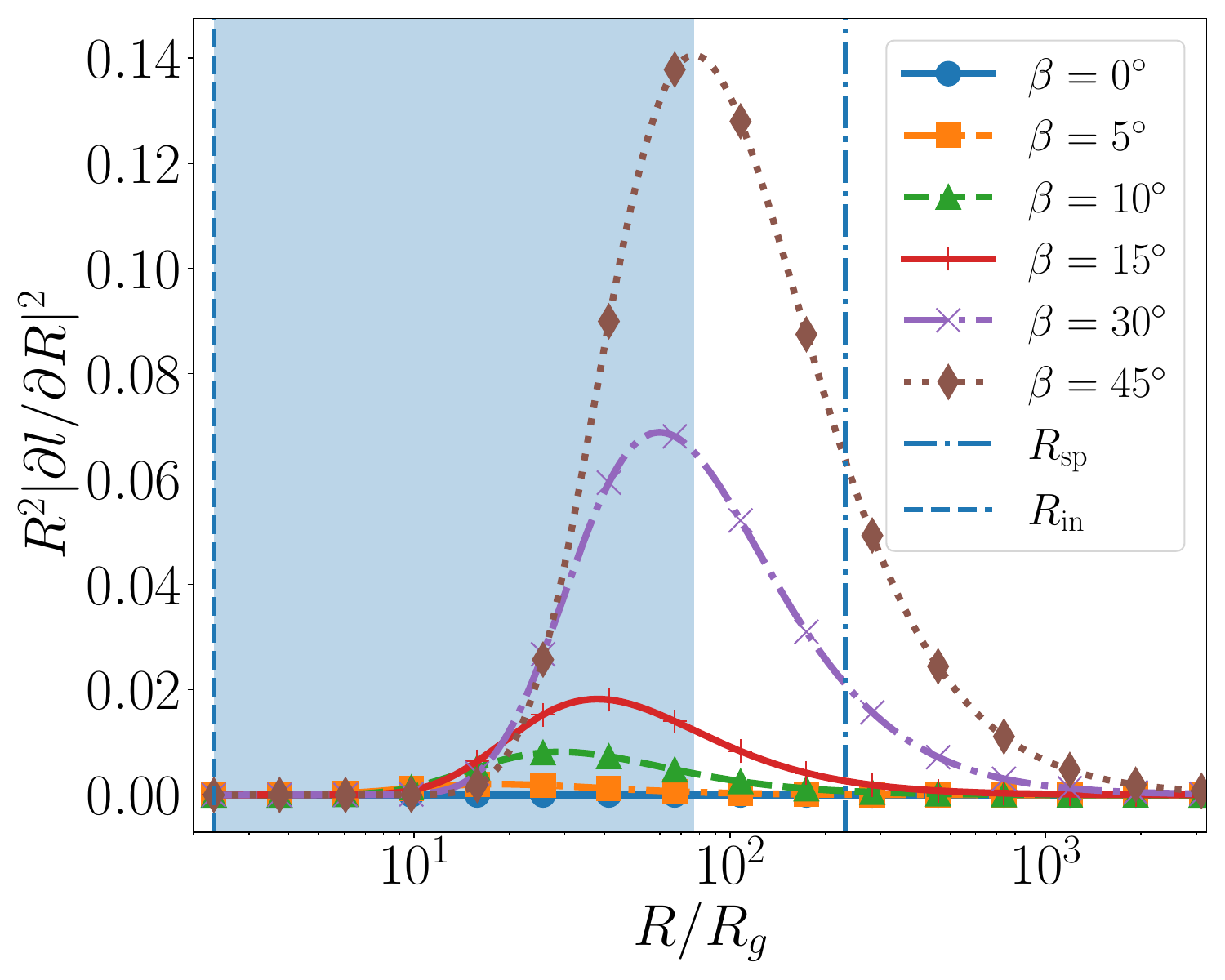}
    \caption{Local disk warp strength as a function of radius when $\alpha = 0.01$. It is shown here that the warp strength increases as the warp angle $\beta$ increases.}
    \label{fig:R2dldr2}
\end{figure}

It should also be noted that the more important quantity is $R^2 | \partial \vec{l}/\partial R|^2$ (shown in Figure \ref{fig:R2dldr2}) because it directly appears in multiple different formulae (as shown in the next sections). Physically, this quantity represents the local strength of the disk warp. The factor of $R^2$ is included to make quantity dimensionless.

Throughout the paper, we refer to the radial interval between the first non-zero value of $R^2 \left| \partial \vec{l} / \partial R \right|^2$ and maximum value of $R^2 \left| \partial \vec{l} / \partial R \right|^2$ as the ``warp region.'' This is a convenient shorthand because it is within this interval that the warp-related modifications to the disk are concentrated.

\subsection{Deriving the mass accretion rate}

To find the mass accretion rate as a function of radius through the disc, we solved the conservation equations above for $V_{\rm R}$ to obtain:
\begin{equation}
    \label{Infall-Velocity}
    V_R = \frac{\frac{\partial}{\partial R}\left(\nu_1\Sigma R^3 \frac{\partial \Omega}{\partial R}\right) - \frac{1}{2}\nu_2\Sigma R^3 \Omega\left|\frac{\partial \vec{l}}{\partial R}\right|^{2}}{\Sigma R \frac{\partial}{\partial R}\left(R^{2} \Omega\right)} ~.
\end{equation}
which the full derivation is described in Appendix \ref{appendix:B}.

Since $\dot{M}_{\rm in}(R) \equiv - 2 \pi R \Sigma V_{\rm R}$, where $\dot{M}(R)$ is the local mass inflow rate at radius $R$, we can rearrange this equation to get:
\begin{equation}
\label{eq:dotMR}
    \dot{M}(R) = 2 \pi \left[3 R^{1/2} \frac{\partial}{\partial R}\left(\nu_1 \Sigma R^{1/2}\right) + \nu_2 \Sigma R^2 \left|\frac{\partial \vec{l}}{\partial R} \right|^2 \right] ~
\end{equation}
where we can also say that $\dot{M}(R) = \dot{M}_0 - \dot{M}_{\rm wind}(R)$ where $\dot{M}_0$ is the accretion rate at the outermost boundary and $\dot{M}_{\mathrm{wind}}(R)$ is the local mass loss due to winds.

\subsection{Defining $R_{\rm in}$ and $R_{\rm sp}$}

For the innermost boundary we define $R_{\rm in}$ as
\begin{equation}
    R_{\rm in} = R_{\rm ISCO} = 3 + z_2 - \sqrt{(3 - z_1)(3 + z_1 + 2z_2)}
\end{equation}
where $z_1 = 1 + (1 - a^2)^{1/3} \left((1 + a)^{1/3} + (1 - a)^{1/3}
\right)$ and $z_2 = \sqrt{3a^2 + z_1^2}$
and we have defined the spherization radius $R_{\rm sp}$ as
\begin{equation}
    R_{\rm sp} = R_{\rm in} \frac{\dot{M}_0}{\dot{M}_{\rm Edd}} ~.
\end{equation}
where $\dot{M}_{\mathrm{Edd}}$ is the Eddington accretion rate.

\subsection{Surface Density}

There exists a formal solution for the surface density profile, which we present in Appendix \ref{appendA}. However, we are able to define a dimensionless function that gives us the surface density profiles and which removes the need to perform numerical integration each time. This is given by $f(r)$:
\begin{equation}
    \label{fr}
    f(R) = \frac{1}{2 
    R^{1/2}} \int_{R_{ \rm in}}^{R} \frac{\dot{M}_{\rm in}(R^{\prime})}{\dot{M}_0} \frac{I(R_{\rm in},R^{\prime})}{\sqrt{R^{\prime}}} dR^{\prime} ~.
\end{equation}
 This can then be used to define the surface density as
\begin{equation}
    \label{Sigma}
    \Sigma(R) = \frac{\dot{M}_0}{3 \pi \nu_1} f(R) ~,
\end{equation}
where $\nu_1 = \alpha H^2 \Omega_K$.

\subsection{Misalignment Intensity Function}

Because the warp enters the accretion rate equation through the radially varying quantity $|\partial \vec{l}/ \partial R|^2$, directly substituting the tilt profiles from Equations \ref{lx}-\ref{lz} would require repeatedly evaluating the same radial dependence during each iteration of the disk structure calculation. We therefore define a misalignment intensity function, $I(R, R^\prime)$, that stores the cumulative effect of the warp between two radii:
\begin{equation}
    \label{IofR}
    I(R, R^{\prime}) = \exp \left[ - \int_{R}^{R^\prime}  \left(\frac{\nu_2}{\nu_1}\right) \frac{x}{3} \left| \frac{\partial \vec{l}}{\partial x} \right|^2 dx \right] ~.
\end{equation}
This function cannot be evaluated analytically and is therefore computed
numerically. 

We begin by defining
\begin{equation}
h(R) = \frac{R}{3} \left(\frac{\nu_{2}}{\nu_{1}}\right)
\left|\frac{\partial \vec{l}}{\partial R}\right|^{2} ~,
\end{equation}
where
\begin{equation}
\label{nu2onnu1}
\frac{\nu_{2}}{\nu_{1}} = \frac{1}{2\alpha^{2}}.
\end{equation}
We then calculate the cumulative integral
\begin{equation}
C(R) = \int_{R_{\rm in}}^{R} h(x) dx ~.
\end{equation}
For two radial grid points satisfying
$r^\prime \geq r$, the integral can be rewritten as:
\begin{equation}
    \int_{R}^{R^{\prime}}h(x) dx = C(R^\prime) - C(R) ~.
\end{equation}
Therefore the misalignment intensity function can be written as:
\begin{equation}
I(R,R^\prime)= \exp [-(C(R^\prime)-C(R))] ~.
\end{equation}
This approach allows us to calculate the matrix once and reuse it for each required computation.

\subsection{Energy conservation}
Now that we have derived the mass accretion rate and surface density, the remaining disk structure is determined through the balance between viscous heating, radiative cooling, and advective transport. We begin with the equation for conservation of energy:
\begin{equation}
    Q^+ = Q_{\rm adv} + Q_{\rm rad} ~,
\end{equation}
where $Q^+$ is the total viscous dissipation rate per unit area, $Q_{\rm adv}$ is the energy advected per unit area, and $Q_{\rm rad}$ is the radiative cooling per unit area. $Q_{\rm rad}$ is formally split into two parts, the portion used to drive the wind
\begin{equation}
    Q_{\rm wind} = \epsilon_{\rm wind} Q_{\rm rad}
\end{equation}
and that which is assumed to escape
\begin{equation}
    Q_{\rm esc} = (1 - \epsilon_{\rm wind} )Q_{\rm rad} ~,
\end{equation}
where $\epsilon_{\rm wind}$ is the fraction of energy used to accelerate the outflow, following the formulae in \citet{Poutanen_07}.

We use the following expressions for advective and radiative cooling:
\begin{equation}
    \label{Qadv_Qrad}
    Q_{\rm adv}(R) =\frac{4\dot{M}(R) a^2 T_{\rm rad}^8 \xi}{27\pi R^2\Sigma^2 \Omega_K^2} ~
\end{equation}
and
\begin{equation}
    \label{eq:Qrad}
    Q_{\rm rad}(R) = \frac{4acT_{\rm rad}^4}{3\kappa_{\rm s}\Sigma} ~
\end{equation}

For viscous heating we use the following equation:
\begin{equation}
    \label{Qplus}
     Q^{+} (R) = \nu_1 \Sigma R^2 \left(\frac{\partial \Omega}{\partial R}\right)^2+\frac{1}{2}\nu_2\Sigma
R^2\Omega^2\left| \frac{\partial{\vec{l}}}{\partial R}\right|^2 ~,
\end{equation}
where the first term is the heating due to differential rotation and the second term is the heating due to the warp.

Using these three equations, we can solve for the midplane radiative temperature:
\begin{equation}
    T_{\rm rad} = \left( \frac{Q^+}{T^4 C_{\rm adv} + C_{\rm rad}} \right)^{1/4} ~,
\end{equation}
where $C_{\rm adv} = 4\dot{M}(R) a^2 T_{\rm rad}^4 \xi/(27\pi R^2\Sigma^2 \Omega_K^2)$ and $C_{\rm rad} = 4ac/(3 \kappa_s \Sigma)$. Equation \ref{eq:Qrad} is the radiative flux equation in \citet{Lipunova_99} and Equation \ref{Qadv_Qrad} is taken directly from \citet{Abramowicz95}.

The advective parameter, $\xi$, is defined as:
\begin{equation}
    \xi = -24 \frac{d\ln(T)}{d\ln R} + 7 \frac{d \ln (\Sigma)}{d \ln R} -9
    \label{xi}
\end{equation}
and is of the order unity. As in Equation \ref{Qadv_Qrad}, this is taken from \citet{Abramowicz95} for a radiation pressure dominated disk in LTE.

As the temperature equation is implicit through both $Q_{\rm adv}$ and $\xi$, we have to iteratively solve for $\xi$, which we do by recasting and solving the equation as a quadratic in $x=T^4$, finding $\xi(R)$ via equation (\ref{xi}), and then iterating until the residual of $\xi(R)$ approaches zero. Finding the temperature this way allows us to individually determine the profiles for the advective and radiative components of the cooling as well as the scale height of the disk.

%\subsection{Iterating $\dot{M}(r)$}\

Unfortunately, the radial mass accretion rate $\dot{M}(R)$ cannot be solved analytically due to the disk structure, radiative cooling, and wind mass loss being mutually coupled, i.e. the local radiative flux determines the amount of energy available to drive the winds, and the resulting mass loss modifies the accretion rate and therefore the disk structure. To find a self-consistent solution, we solve for $\dot{M}(R)$ by first assuming $\dot{M}(R) = \dot{M}_0 \frac{R}{R_{\rm g}}$ as an initial guess. We then solve the structure for a flat, unwarped disk to find the average scale height, $\left< H_\mathrm{flat} \right> / R$, which then enters equation (\ref{Rwarp}) to determine the warp radius for the given mass accretion rate profile. Using this warp radius we are then able to solve equation (\ref{IofR}). Finally, we use the warp intensity function to solve equations (\ref{fr})-(\ref{Qplus}) to find the local radiative cooling rate $Q_{\rm rad}(R)$.

\subsubsection{Mass loss rates}

The previously calculated radiative energy has the potential to drive disk winds. To calculate the associated mass loss profile, we rearrange equation (12) in \citet{Poutanen_07} and solve for $d\dot{M}(R)$:
\begin{equation}
    \label{dMwind}
    d\dot{M}_{\rm wind}(R) = \epsilon_{\rm wind} Q_{\rm rad} \frac{8\pi R^2}{GM}dR
\end{equation}
We start from the outer boundary where $\dot{M}(R_{\rm out}) = \dot{M}_0$, and update the mass accretion profile by the cumulative mass loss as we move in through the disk. We use \texttt{Scipy}'s root finding algorithm to solve for the residual of the equation:
\begin{equation}
    \dot{M}_{\rm test}(R) - \dot{M}_{\rm update}(R) = \rm residual
\end{equation}
where $\dot{M}_{\rm test}(R)$ is an intermediate variable within the update loop and $\dot{M}_{\rm update}$ is the change being tested. This comes from equation \ref{dMwind}.
When the residual approaches zero, the resulting mass accretion rate profile will simultaneously satisfy the disk structure equations and the wind mass-loss closure. This allows for a fully self-consistent mass profile across the disc and we can use this to re-solve all the disk structure equations for full self-consistency.

Although the decreased angular momentum in \cite{Fragile26} is attributed to standing shocks, we do not explicitly include these shocks in our model. Our model instead uses the viscous warped disk equations as described in \cite{Pringle_92}, where $\nu_2$ describes the viscosity between neighboring annuli with different angular momentum directions. This allows us to create a simplified representation of the angular momentum transport associated with the warped disk. We are still treating the warp here as if all transport is diffusive. Our comparisons with GR-RMHD in Section \ref{sec:BetaVsMdotIn} show that our model is able to qualitatively recover the increase of the accretion rate at the ISCO with misalignment found in the simulations.

\section{Results}
\label{sec:Results}

We solve the above sets of equations for a range of disk misalignment angles ($\beta = 0^\circ - 45^\circ$). Unless otherwise stated, the calculations given in Sections \ref{sec:MassAccretion}-\ref{sec:BetaVsMdotIn} use:
\begin{enumerate}
    \item Spin Parameter: $a = 0.9$
    \item Black Hole Mass: $M = 10 M_{\odot}$
    \item Alpha-Viscosity Parameter: $\alpha = 0.01$
    \item Outer Accretion Rate: $\dot{M}_0 = 100 \dot{M}_{\rm edd}$
    \item Wind Parameter: $\epsilon_{\rm wind} = 1.0$
\end{enumerate}

We have chosen $\epsilon_{\rm wind}=1.0$ because it produces $\dot{M}(R_{\rm in})$ of order Eddington for the unwarped disk. We further explore the effects of $\epsilon_{\rm wind}$ on the disk structure in Section \ref{sec:e_wind}. For each inclination, the iterative solver produces self-consistent radial profiles for the mass accretion rate, mass loss rate, surface density, temperature, scale height, and energy partition. These profiles allow us to determine how disk misalignment affects disk structure and energy transport within the disk. It should be noted that due to some numerical instabilities coming from a combination of $\Sigma(0) = 0$ and the use of numerical derivatives, we utilize a simple smoothing function (a moving average) and force the innermost radial cells to behave in a physically appropriate. manner (i.e., by setting the innermost cell to be the same as the second innermost cell). In addition to the smoothing and normalization, the code imposes small, positive floors on quantities such as $\dot{M}$ and $\Sigma$.

\subsection{Mass Accretion Rates}
\label{sec:MassAccretion}

\begin{figure*}
    \centering    \includegraphics[width=1.0\linewidth]{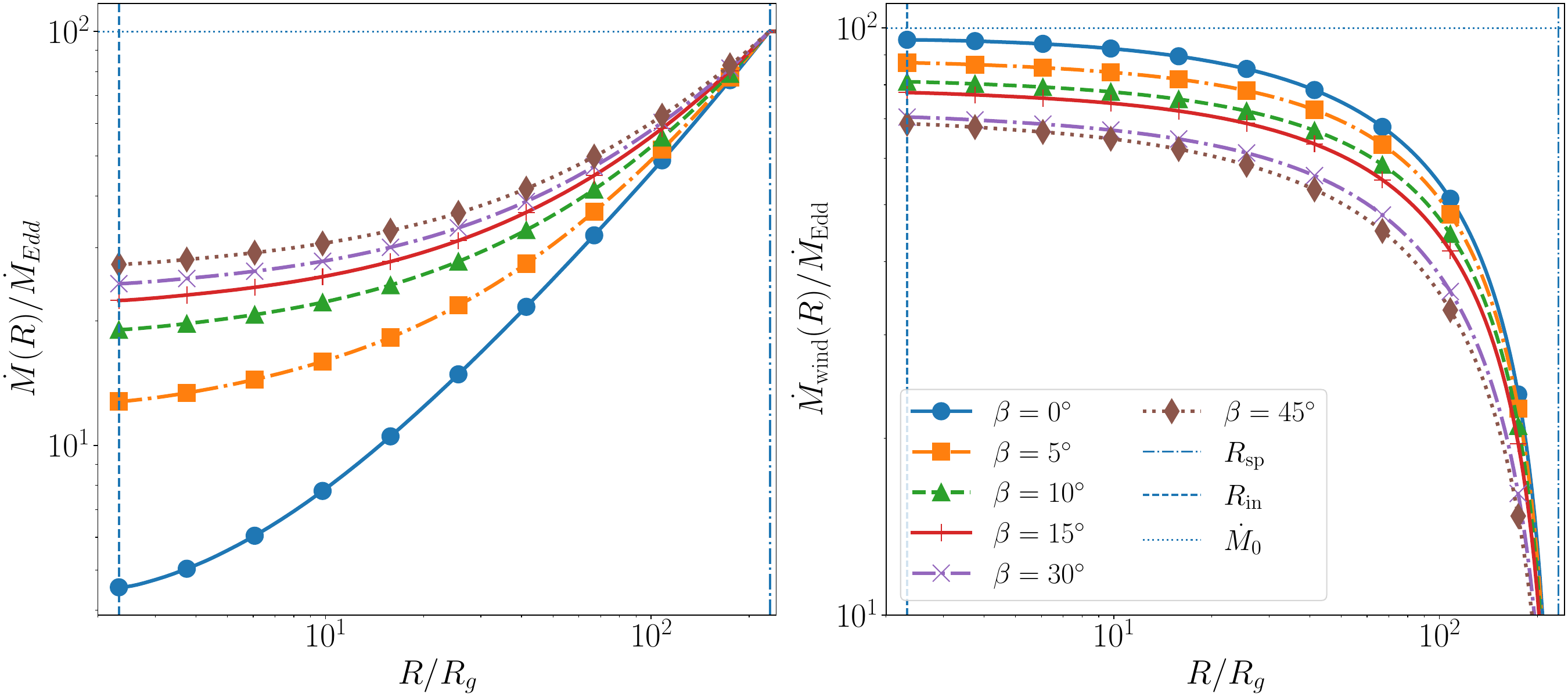}
    \caption{Mass accretion rate (left) and mass loss (right) profiles as a function of radius.}
    \label{fig:MdotR}
\end{figure*}

Figure \ref{fig:MdotR} (left panel) shows the radial profile of the mass accretion rate, normalized such that 1 on the y-axis is the Eddington accretion rate. For the unwarped case, ($\beta = 0^\circ$), we are able to closely reproduce the analytical solution for a super-Eddington accretion disk with mass loss as derived in \citet{Lipunova_99, Poutanen_07}. Specifically, we find that $\dot{M}(R)$ decreases monotonically with decreasing radius, falling from the supplied accretion rate, $\dot{M}_0$, to a value of order Eddington at the ISCO.

For disks with non-zero warp, ($\beta > 0^\circ$), we find that the $\dot{M}(R)$ profiles are dependent on the misalignment angle. We find that, compared to the unwarped case, the warped discs exhibit much higher $\dot{M}(R)$ values at the ISCO (similar to the finding of \cite{Lodato06} at sub-Eddington rates and \cite{Fragile26} in super-Eddington cases) and that the $\dot{M}(R)$ profiles are much flatter than the unwarped case. %In addition to this, we find that all the profiles converge near the spherization radius, suggesting that the warp is more effective closer to the ISCO and that the main effect is to modify how the inflowing mass is distributed.

\subsubsection{Mass loss profiles}

Figure \ref{fig:MdotR} (right panel) shows the cumulative mass lost through winds as a function of radius for various disk warp angles, again normalized such that 1 on the y-axis is the Eddington accretion rate. For the unwarped case, the profile closely follows the expected behavior of the standard super-Eddington disk models, with most of the mass being lost to winds before reaching the ISCO. This mass loss is the reason for the low accretion rate at the ISCO for the unwarped disk.

As the warp angle increases, the total mass lost to the wind decreases significantly. The warped disks with the largest misalignment angles show the least amount of mass loss throughout the inner disk, allowing for a much larger fraction of the supplied material to reach smaller radii (and explains Figure \ref{fig:MdotR} (left panel), where the strongest warps show the highest $\dot{M}(R)$ values at the ISCO).

The reduction in wind mass loss can be explained as a direct consequence of the increasing dominance of advection over radiation as the primary cooling mechanism within the inner disk (see following section). Due to the cumulative mass loss being directly proportional to $Q_{\rm rad}$ (equation \ref{dMwind}), as the radiative energy decreases, so does the amount of mass lost via the wind.

The convergence of all profiles as the radius approaches the spherization radius indicates that the effects of the warp are confined to the inner disk; outside of the spherization radius, the disk structure is largely independent of the warp inclination, which is the expected behavior resulting from our assumptions that the warp primarily modifies the inner disk.

The enhancement of inward mass transport previously found for warped disks at sub-Eddington accretion rates therefore also persists into the super-Eddington regime considered here. In the present models, however, this effect is coupled to wind launching: stronger warps favor advection over radiation, reducing the radiative energy available to drive the wind and allowing a larger fraction of the supplied material to reach the ISCO.

\subsection{Surface Density Profiles}

\begin{figure}
    \centering
    \includegraphics[width=1.0\linewidth]{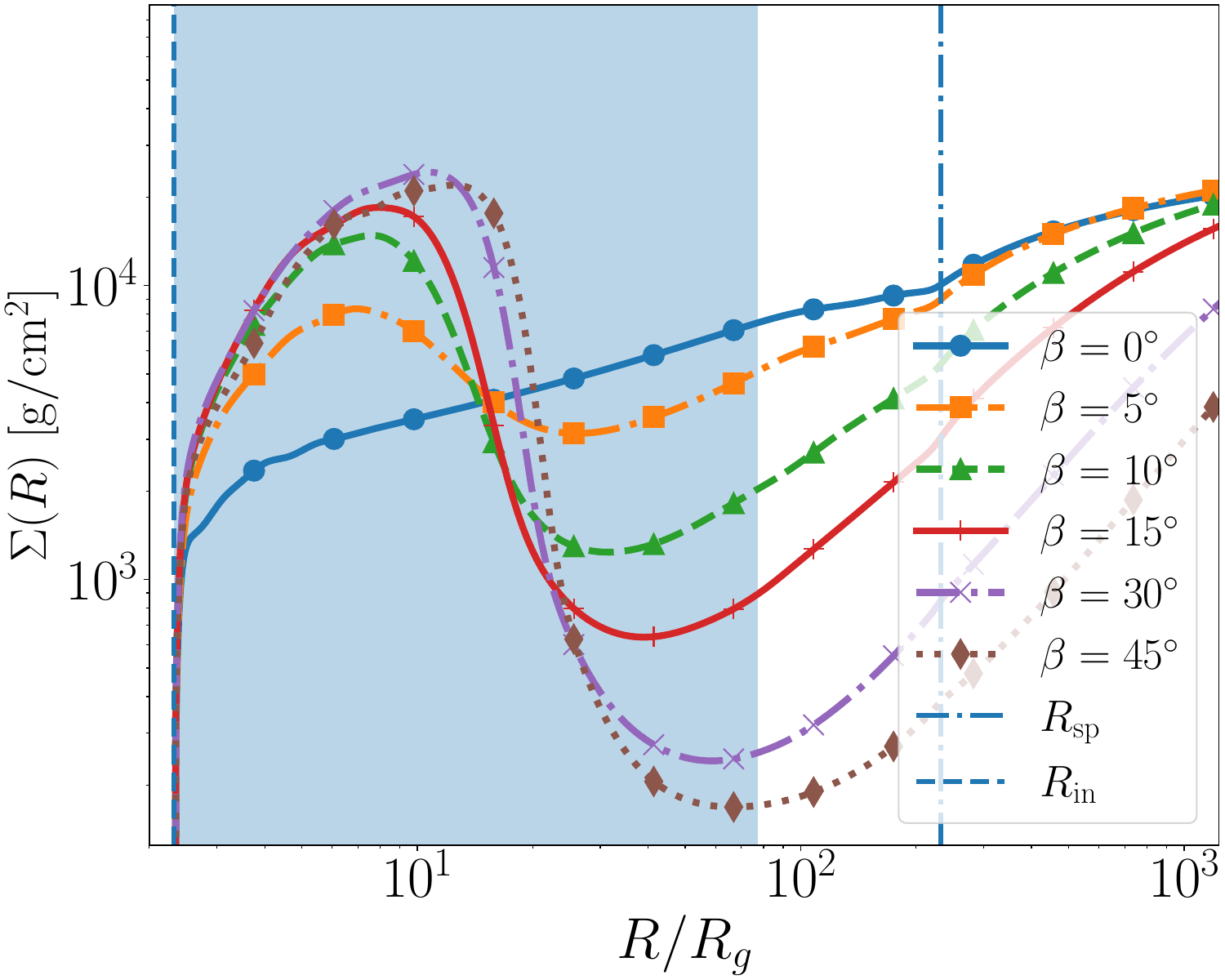}
    \caption{Surface density as a function of radius. We can see that much of the modification to the profile happens within the blue shaded warp region.}
    \label{fig:sigma}
\end{figure}

Figure \ref{fig:sigma} shows that the surface density changes drastically once a warp is introduced. In the unwarped case, the surface density decreases monotonically from the outer boundary to the ISCO \citep[matching that derived by][]{Pringle_92}. In contrast, for the warped discs, the surface density is not monotonic. Using an inside-out point of view, it reaches a minimum within or just beyond the warp region and then increases as it approaches the spherization radius. This behavior indicates that the warp changes the distribution of mass within the disk rather than simply rescaling the unwarped solution. As the strength of the warp increases, the inner peak increases in strength while the minimum becomes flatter and more extended. At large radii, however, all profiles converge beyond the spherization radius, indicating that the effect of the warp is concentrated in the inner disk. We also find that the profile approaches that of the unwarped disk asymptotically as $\beta \xrightarrow{}0$.

Since the surface density is recovered through equation (\ref{Sigma}), its radial profile depends on both the misalignment intensity function $I(R)$ and the local mass accretion rate. This demonstrates the coupling between the warp, wind-driven mass loss, and the redistribution of mass within the disk, explaining why the strongest changes in surface density occur within the warp region.

\subsection{Energy Profiles}

\begin{figure*}
    \centering
    \includegraphics[width=1.0\linewidth]{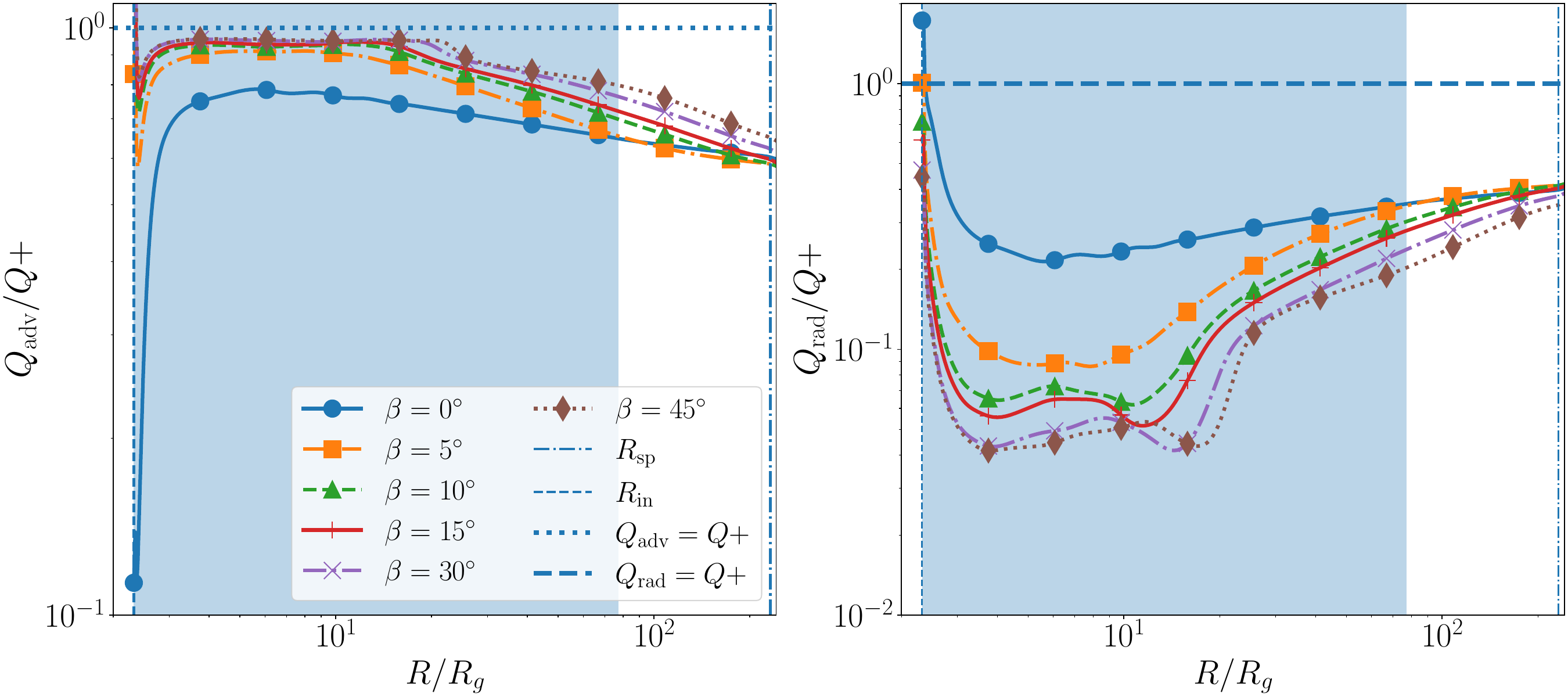}
    \caption{Plots of the ratio of advective cooling (left) and radiative cooling (right) to total heating as a function of radius. For highly warped disks, we find that, within the spherization radius,  $Q_{\rm adv} \sim Q+$, which is the result that we were expecting from \citet{Lipunova_99}.}
    \label{fig:advplus&radplus}
\end{figure*}

\begin{figure}
    \centering
    \includegraphics[width=1.0\linewidth]{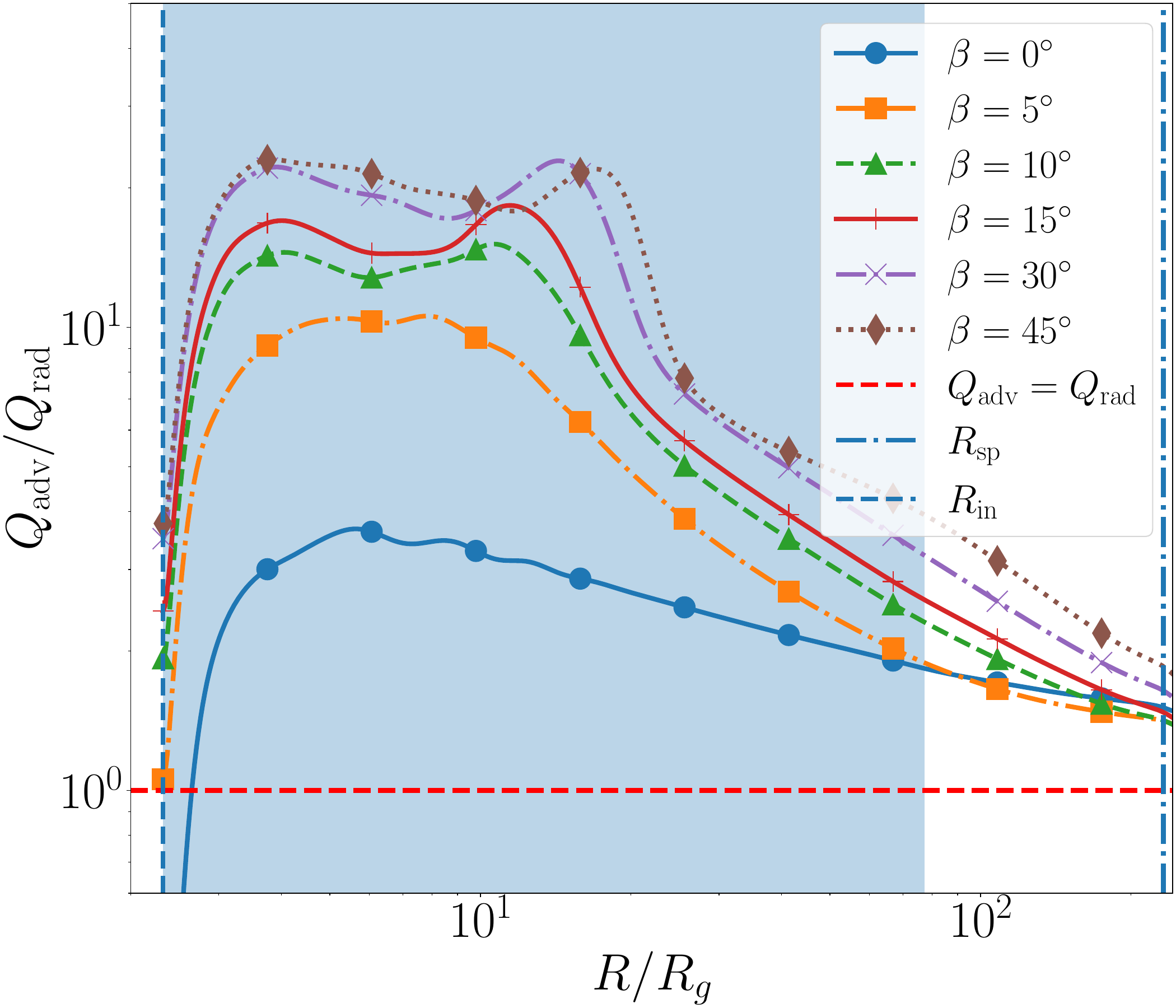}
    \caption{Plot of the ratio of advective cooling to radiative cooling versus radius. The graph shows clear advection domination within the spherization radius.}
    \label{fig:advrad}
\end{figure}

Figures \ref{fig:advplus&radplus} and \ref{fig:advrad} show that a warp strongly affects the energy distribution within the inner disk. In agreement with numerical simulations \citep{Fragile26}, we find that in the unwarped case, advection is still the dominant cooling mechanism, although radiation plays a larger role than in the warped cases. This can be seen in Figure \ref{fig:advrad}, where the ratio of $Q_{\rm adv} / Q_{\rm rad}$ is on the order of unity.

In contrast, once a non-zero warp angle is introduced, the inner disk becomes significantly more advective. The ratio of $Q_{\rm adv} / Q_{\rm rad}$ exceeds unity and reaches its peak within the warp region, indicating that advective transport dominates over radiative cooling within that region. This same trend is seen in the plots of $Q_{\rm adv} / Q+$, where it approaches unity within the spherization radius for all disks with large non-zero misalignment angle, indicating that most of the energy is being advected inwards, rather than being radiated away.

For large non-zero warps, $Q_{\rm rad} / Q+$ appears greatest close to the outer boundary before sharply decreasing within the warp region. This decrease shows that the warp-driven redistribution of the energy partition is concentrated primarily within the inner disk, where the additional warp-related dissipation and the altered disk structure favor advective transport over radiative cooling. We use the term radiative cooling; however, because $\epsilon_{\rm wind} = 1.0$, all the radiative energy is being used to drive winds, so a more accurate description might be wind cooling.

\subsection{Disk Scale Height}

\begin{figure}
    \centering
    \includegraphics[width=1.0\linewidth]{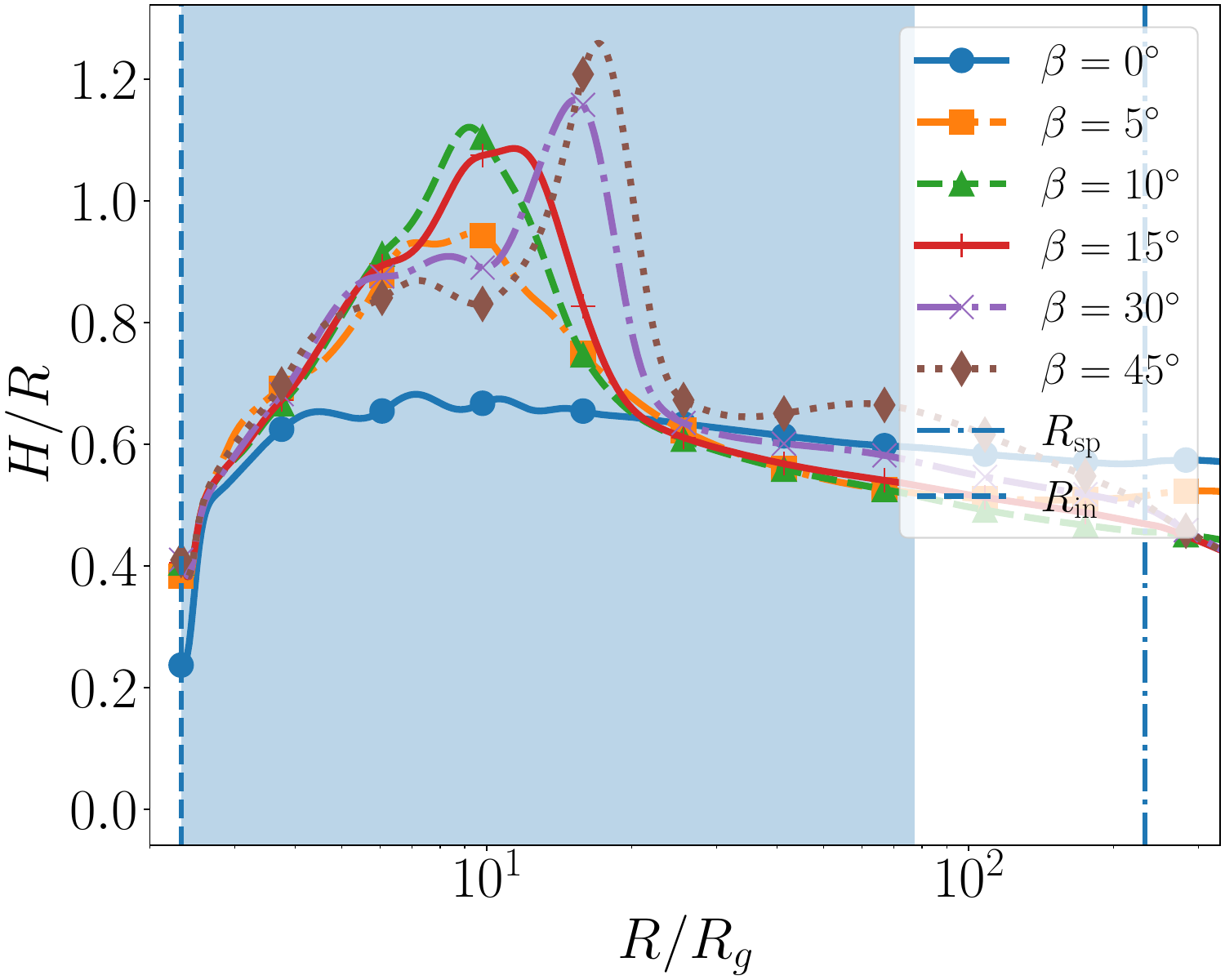}
    \caption{Plot of scale height as a function of radius.}
    \label{fig:HonR}
\end{figure}

Figure \ref{fig:HonR} presents the scale height of the disk, expressed as a dimensionless ratio of $H / R$. Our results for an unwarped disk match those found in both \citet{Lipunova_99} and \citet{Poutanen_07}. In contrast, the warped discs show an increase in scale-height of the inner disk, peaking at $\sim 10 R_g$. The scale-height then decreases as one moves outward through the disk, finally converging around the same height as the unwarped disk near the spherization radius. 

The maximum scale-height increases with the misalignment angle of the warp, demonstrating that a stronger warp produces a more vertically extended inner disk. For the largest misalignment values, the pressure scale height exceeds $H/R \sim 1$, possibly indicating additional mass loss at those radii that is unaccounted for in Figure \ref{fig:MdotR}. If such mass loss occurs, inward mass accretion and surface density would decrease thereby decreasing the advective fraction. This means that for highly misaligned disks, our model may be overestimating the advective fraction as well as the amount of material reaching the ISCO.

\subsection{Exploring additional parameters}

The previous sections used the parameters stated at the beginning of Section \ref{sec:Results}. We now vary $a$, $\alpha$, and $\epsilon_{\rm wind}$ individually while holding the remaining parameters fixed, allowing their separate effects on the disk structure and energy partition to be identified. 

\subsubsection{Decreasing spin parameter $a$}
\label{sec:a}

Decreasing the spin parameter $a$ from $a=0.9$ to 0.5 shifts the ISCO outward while also shifting the warp radius inward. This results in radially compressed profiles when compared to Figures \ref{fig:MdotR} - \ref{fig:HonR}. We choose $a=0.5$ as the lower comparison value because, for smaller spins, the predicted warp radius approaches or falls inside the ISCO, leaving no radially resolved warped region within the model.

\begin{figure*}
    \centering
    \includegraphics[width=0.49\linewidth]{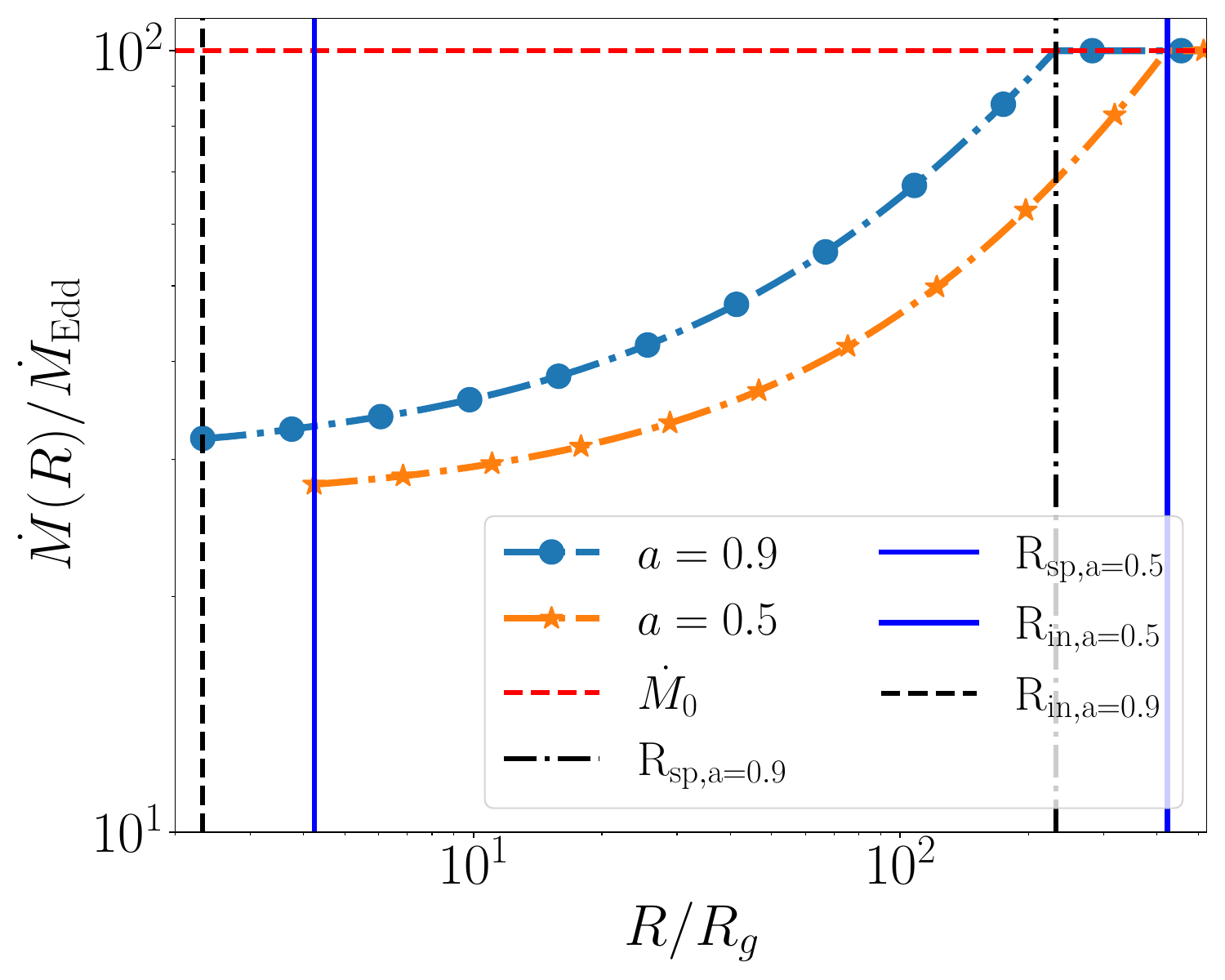}
    \includegraphics[width=0.49\linewidth]{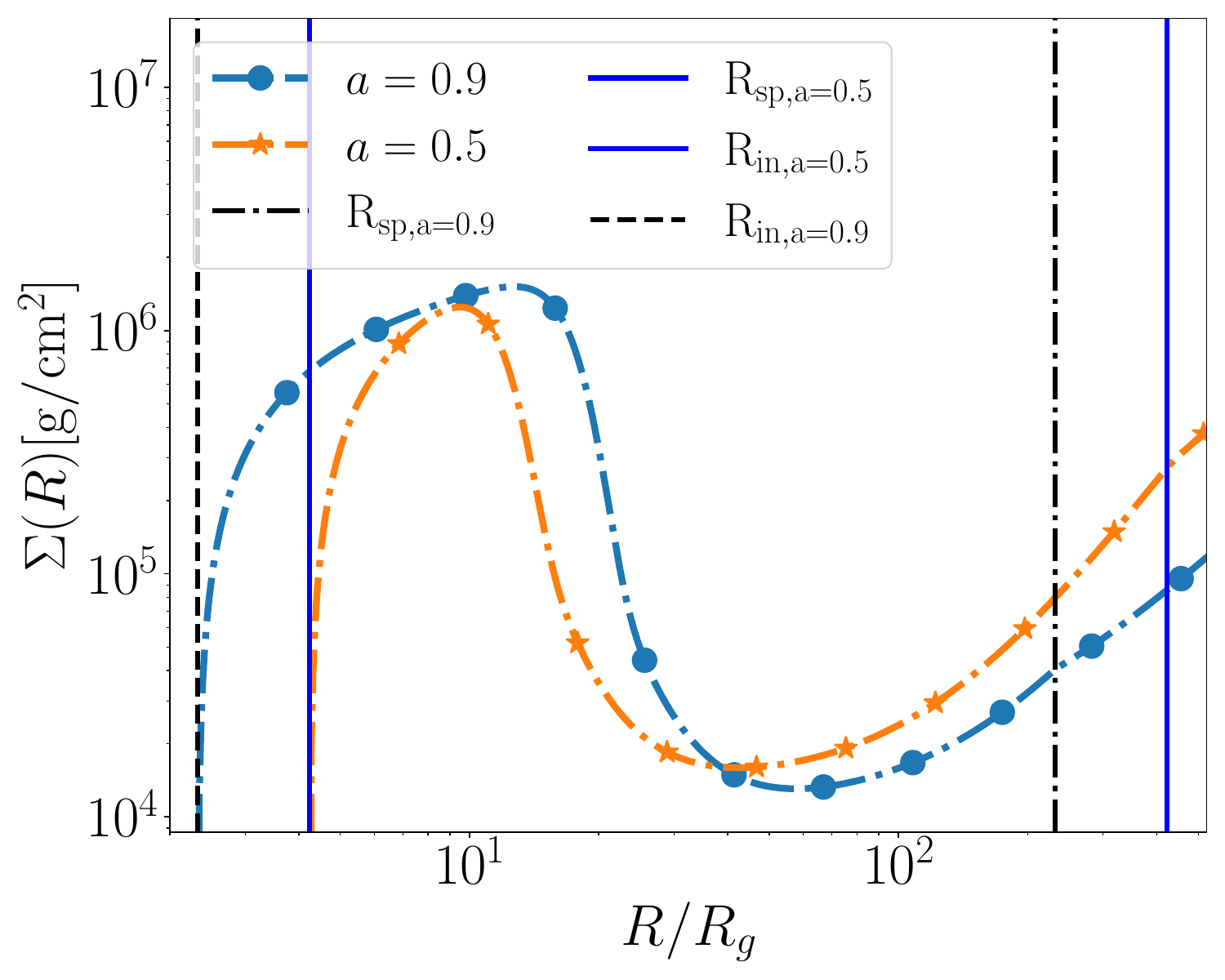}
    \caption{Mass accretion rate (left) and surface density (right) as a function of radius for $a =0.5$ and $a=0.9$ (both with $\beta = 45^\circ$).}
    \label{fig:a_0.5_MdotR_etc}
\end{figure*}

\begin{figure*}
    \centering
    \includegraphics[width=0.49\linewidth]{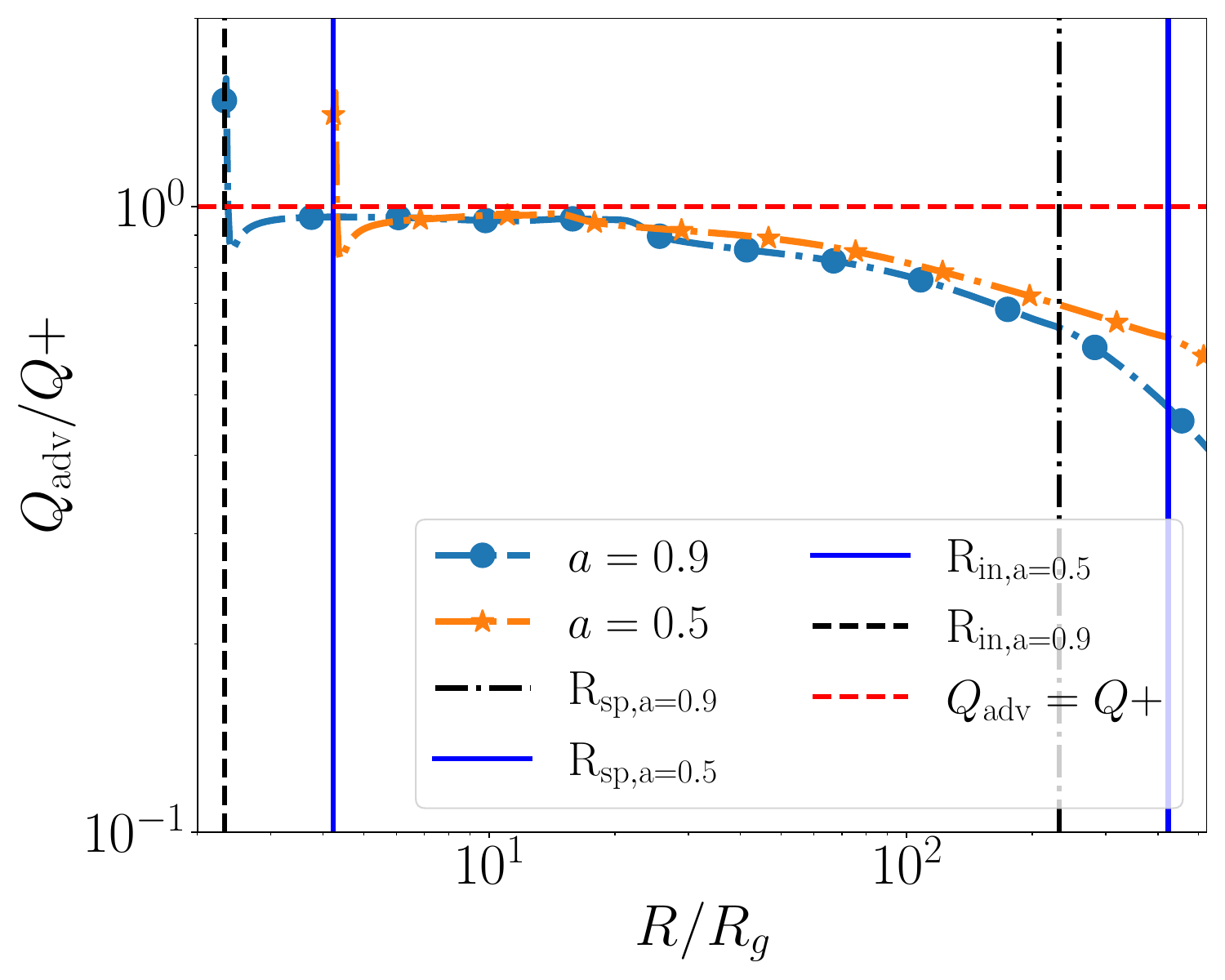}
    \includegraphics[width=0.49\linewidth]{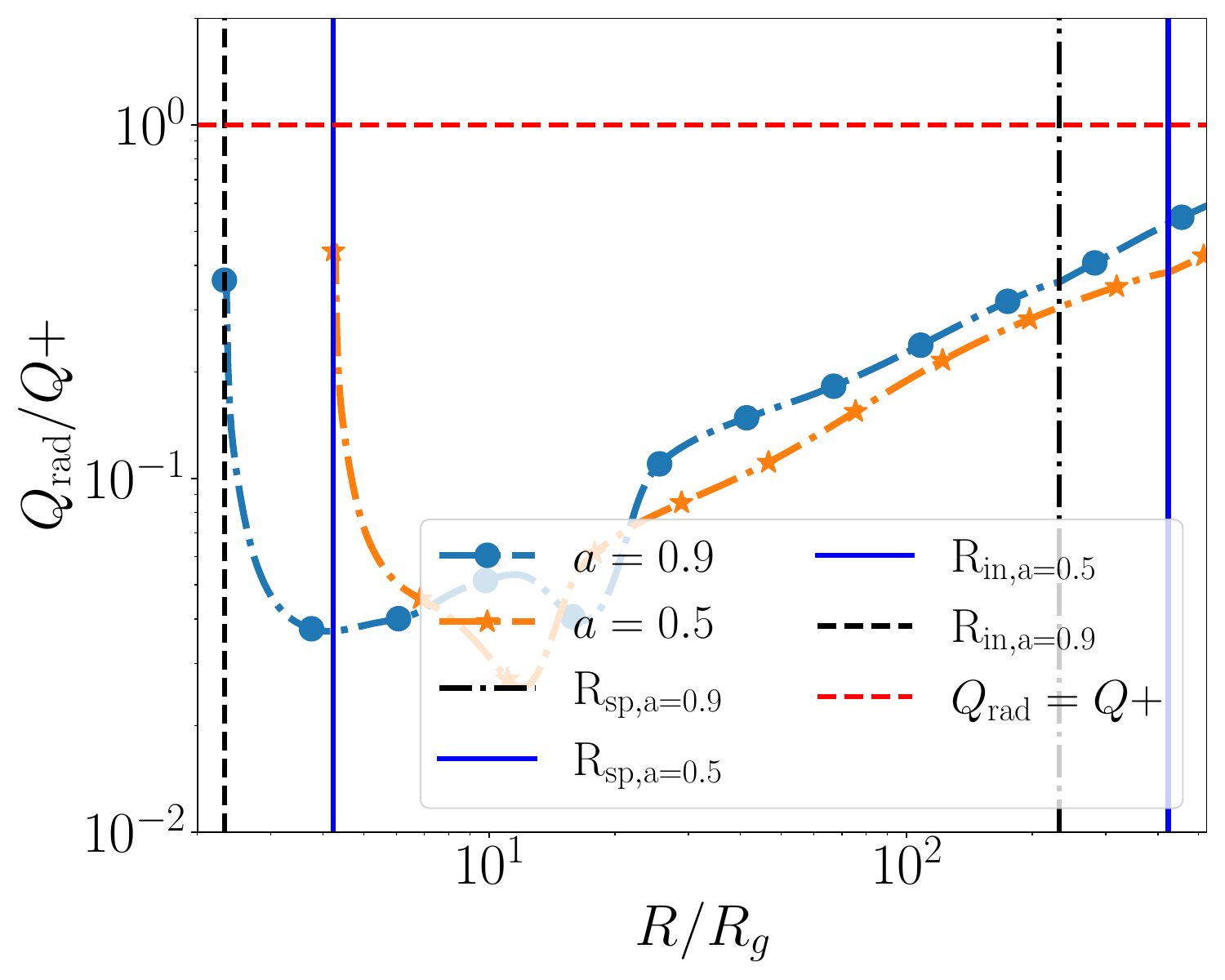}
    \caption{Plots of the ratio of advective cooling (left) and radiative cooling (right) to total heating as a function of radius for $a =0.5$ and $a=0.9$ (both with $\beta = 45^\circ$). }
    \label{fig:a_0.5_etc}    
\end{figure*}

From equation (\ref{Rwarp}), we see that the warp radius scales as $\propto a^{2/3}$. This is a result of the Lense-Thirring torque's dependence upon the spin of the black hole and means that the warp radius moves inwards as the spin decreases. This is clearly visible in Figure \ref{fig:a_0.5_MdotR_etc} (right panel) where the surface-density peak for $a=0.5$ occurs at a smaller radius, while the inner edge of the disk lies at a larger radius because of the outward shift of the ISCO. The energy profiles in Figure \ref{fig:a_0.5_etc} show that the disk remains strongly advective for this range of spin values. 

\subsubsection{Increasing $\alpha$}
\label{sec:alpha}

\begin{figure*}
    \centering
    \includegraphics[width=0.49\linewidth]{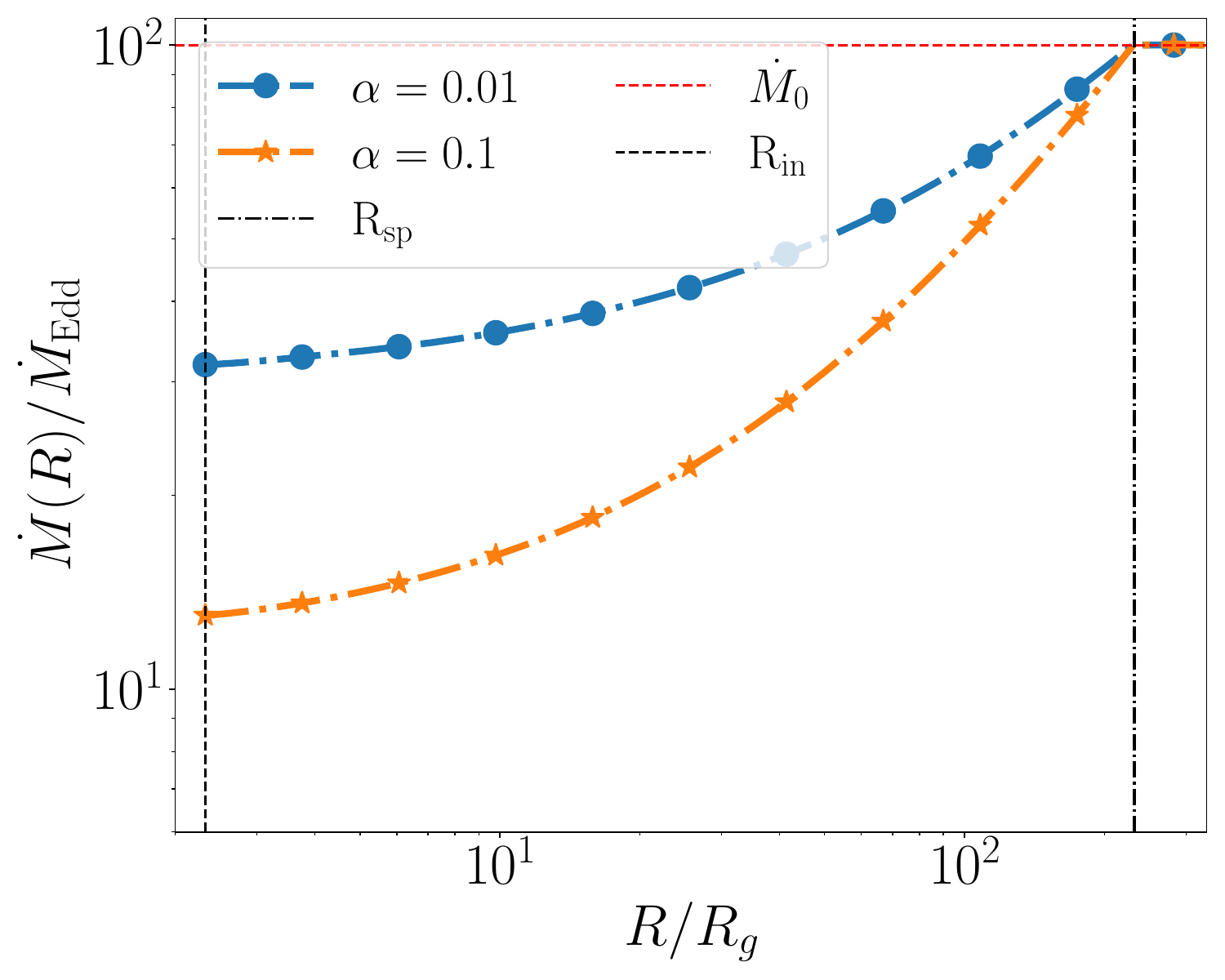}
    \includegraphics[width=0.49\linewidth]{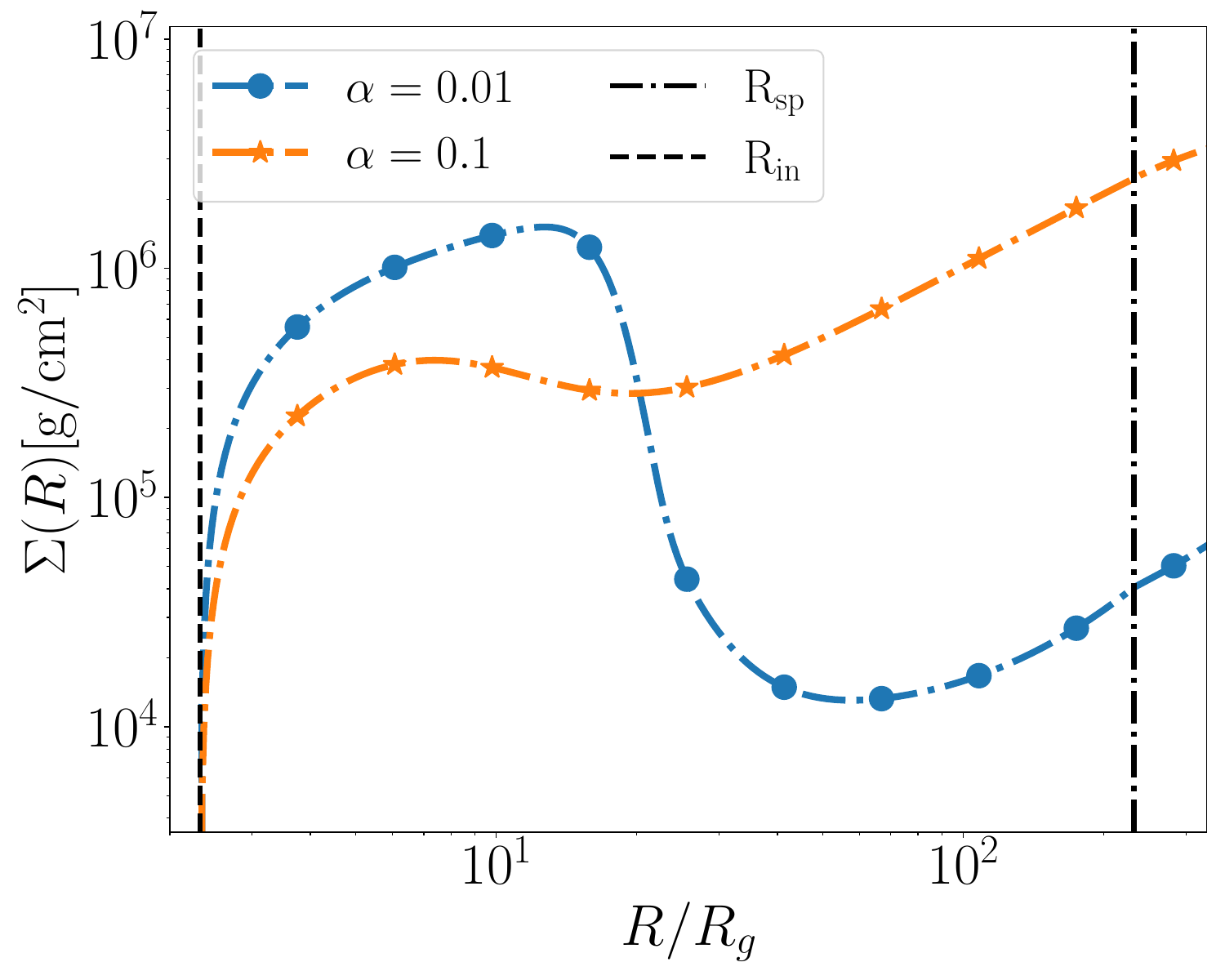}

    \caption{Mass accretion rate (left) and surface density (right) as a function of radius for $\alpha = 0.01$ and $\alpha =0.1$ (both with $\beta = 45^\circ$). As in Figure \ref{fig:sigma}, many of the modifications to the profiles occur within the warp region.}
    \label{fig:alpha_0.1_MdotR_Sigma}
\end{figure*}

Increasing $\alpha$ from $0.01$ to $0.1$ weakens the influence of the warp. From equation (\ref{nu2onnu1}), $\nu_2 / \nu_1 \propto \alpha^{-2}$ so increasing $\alpha$ reduces the relative vertical-shear viscosity responsible for warp-driven angular momentum transport. In Figure \ref{fig:alpha_0.1_MdotR_Sigma} (left panel), the model using $\alpha = 0.1,~\beta = 45^\circ$ reaches an accretion rate of only approximately $13 \dot{M}_{\rm Edd}$ whereas the model using $\alpha = 0.01,~\beta = 45^\circ$ reaches approximately $30 \dot{M}_{\rm Edd}$.

\begin{figure*}
    \centering
    \includegraphics[width=0.49\linewidth]{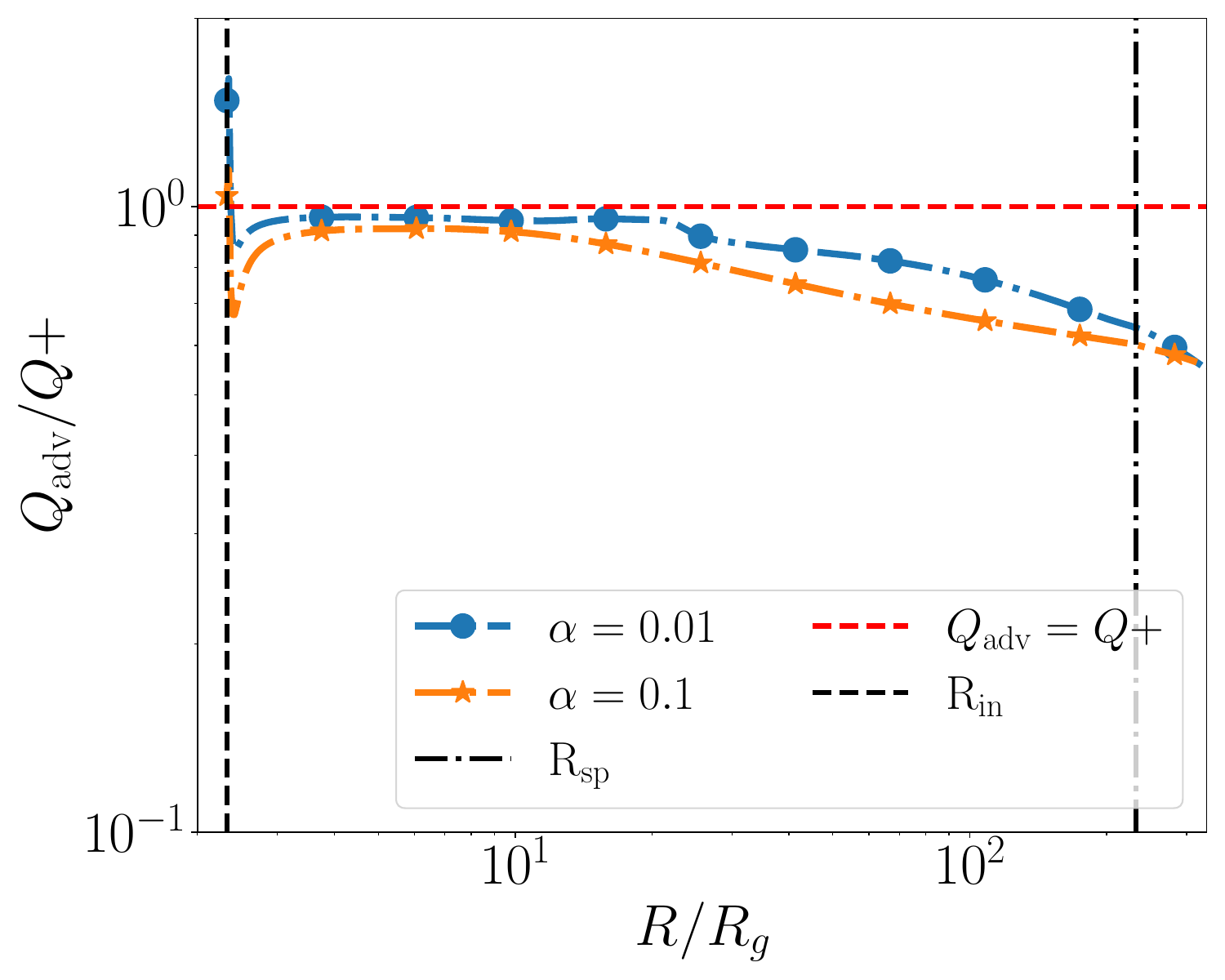}
    \includegraphics[width=0.49\linewidth]{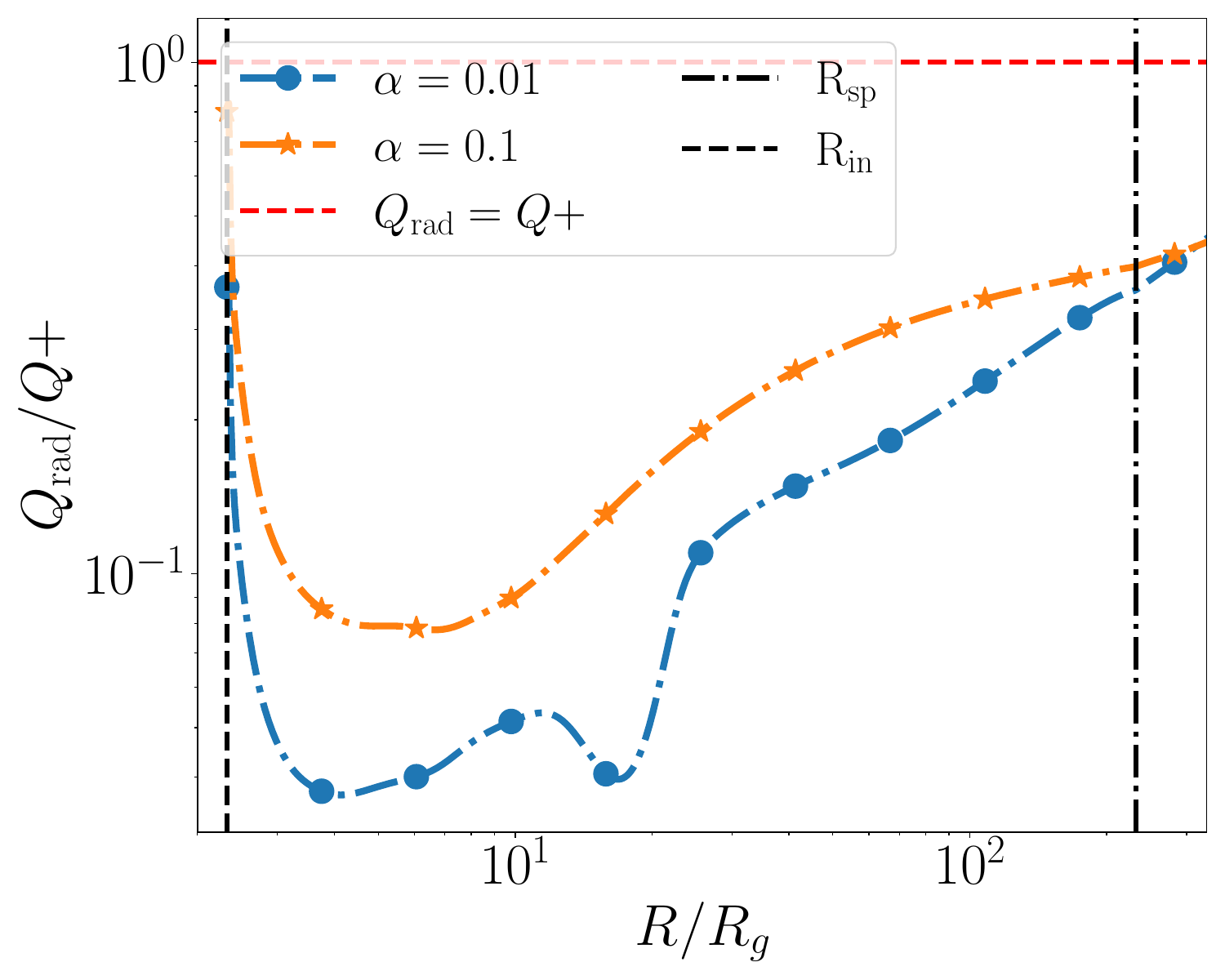}

    \caption{Plots of the ratio of advective cooling (left) and radiative cooling (right) to total heating as a function of radius for $\alpha = 0.01$ and $\alpha =0.1$ (both with $\beta = 45^\circ$). }
    \label{fig:alpha_0.1_advplus_radplus}
\end{figure*}

The reduced influence of the warp is also demonstrated through the surface-density profiles in Figure \ref{fig:alpha_0.1_MdotR_Sigma} (right panel). Increasing $\alpha$ suppresses the large inner surface-density peak, while reaching a considerably larger value at the spherization radius. The energy profiles in Figure \ref{fig:alpha_0.1_advplus_radplus} show the same trend. For $\alpha = 0.1$, the ratio of $Q_{\rm rad} / Q+$ is larger, whereas the ratio of $Q_{\rm adv} / Q+$ is larger in the model using $\alpha = 0.01$.

\subsubsection{Decreasing $\epsilon_{\rm wind}$}
\label{sec:e_wind}

\begin{figure*}
    \centering
    \includegraphics[width=0.49\linewidth]{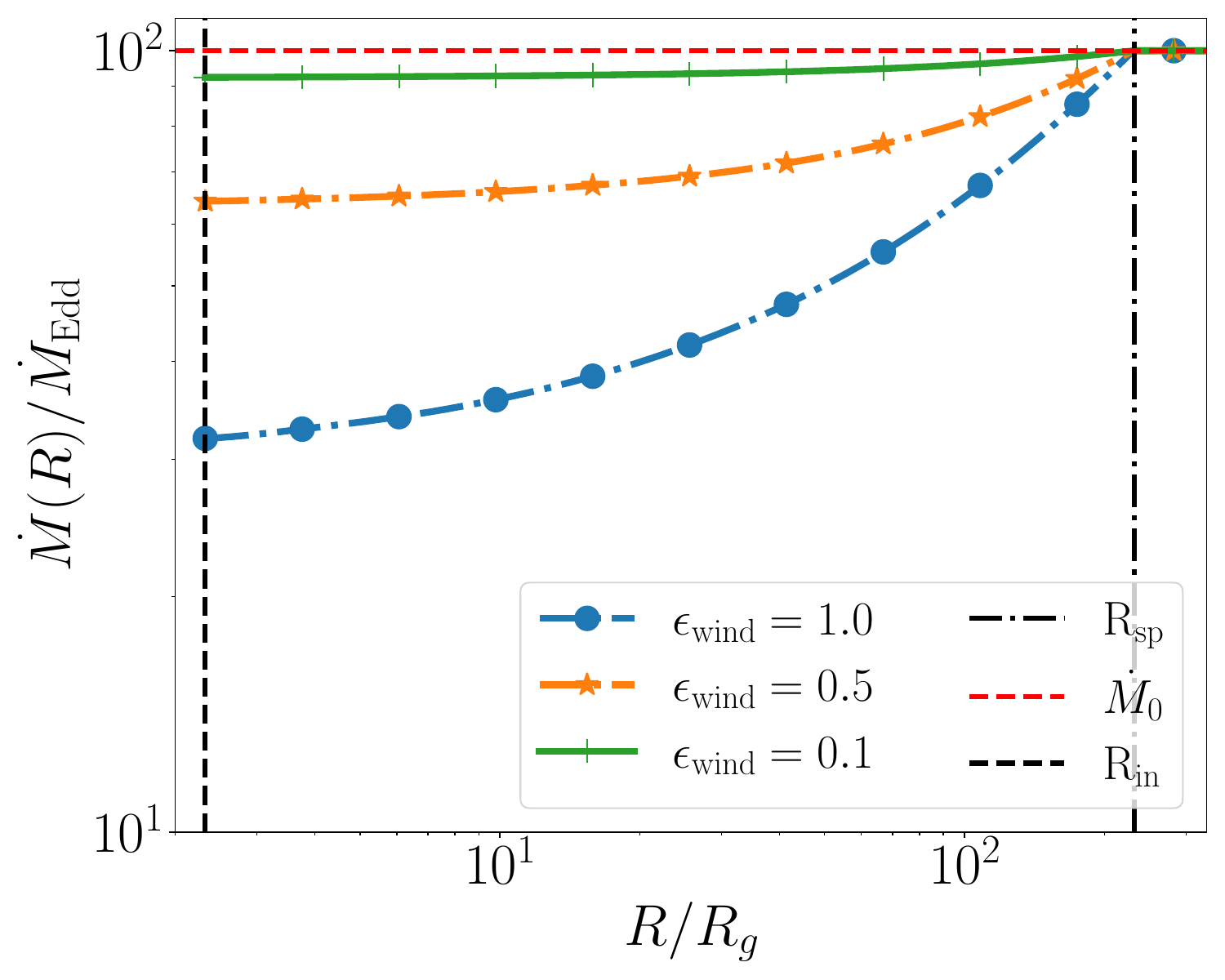}
    \includegraphics[width=0.49\linewidth]{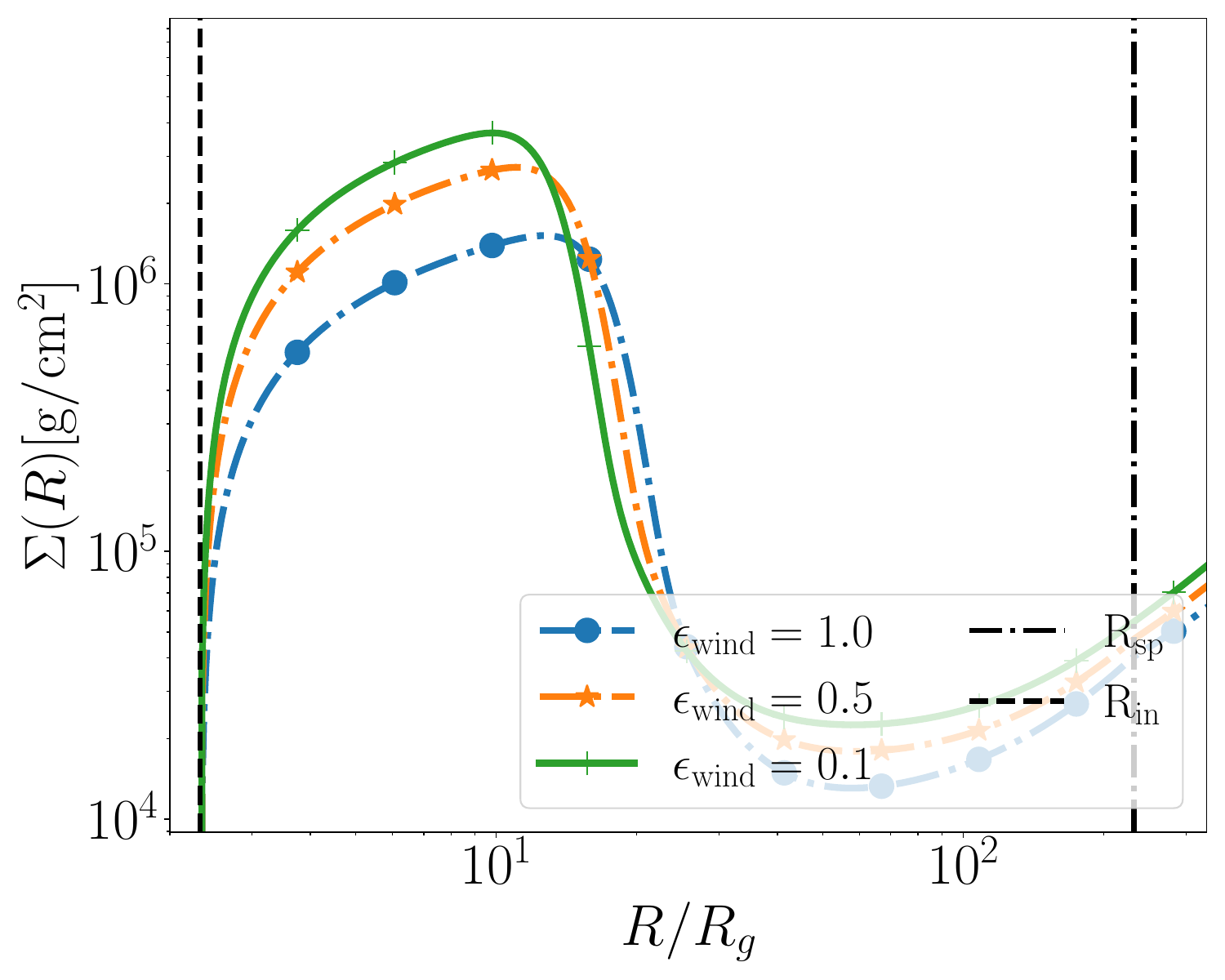}

    \caption{Mass accretion rate (left) and surface density (right) as a function of radius for $\epsilon_{\rm wind} = 0.1, 0.5, 1.0$ with $\beta = 45^{\circ}$.}
    \label{fig:changing_ewind_1}
\end{figure*}

Decreasing $\epsilon_{\rm wind}$ reduces the fraction of the radiative energy used to drive the wind. Since the local mass loss rate is directly proportional to $\epsilon_{\rm wind} Q_{\rm rad}$ through equation (\ref{dMwind}), a smaller value of $\epsilon_{\rm wind}$ results in less mass being removed from the disk. This can be clearly seen in Figure \ref{fig:changing_ewind_1} (left panel), where reducing $\epsilon_{\rm wind}$ from a value of 1.0 to 0.1 flattens the $\dot{M}(R)$ profile and allows a much larger fraction of the initial supplied mass to reach the ISCO.

\begin{figure*}
    \centering
    \includegraphics[width=0.49\linewidth]{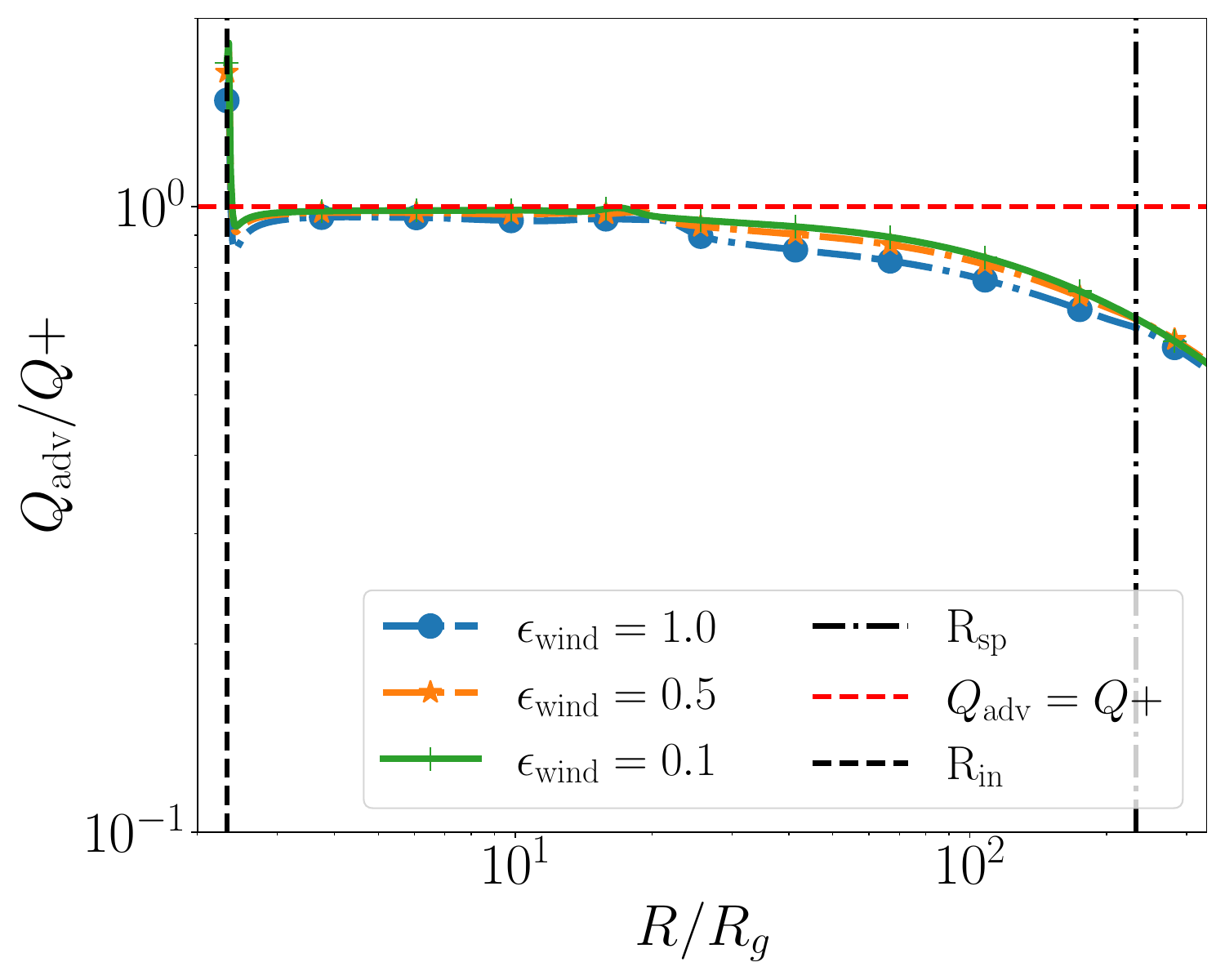}
    \includegraphics[width=0.49\linewidth]{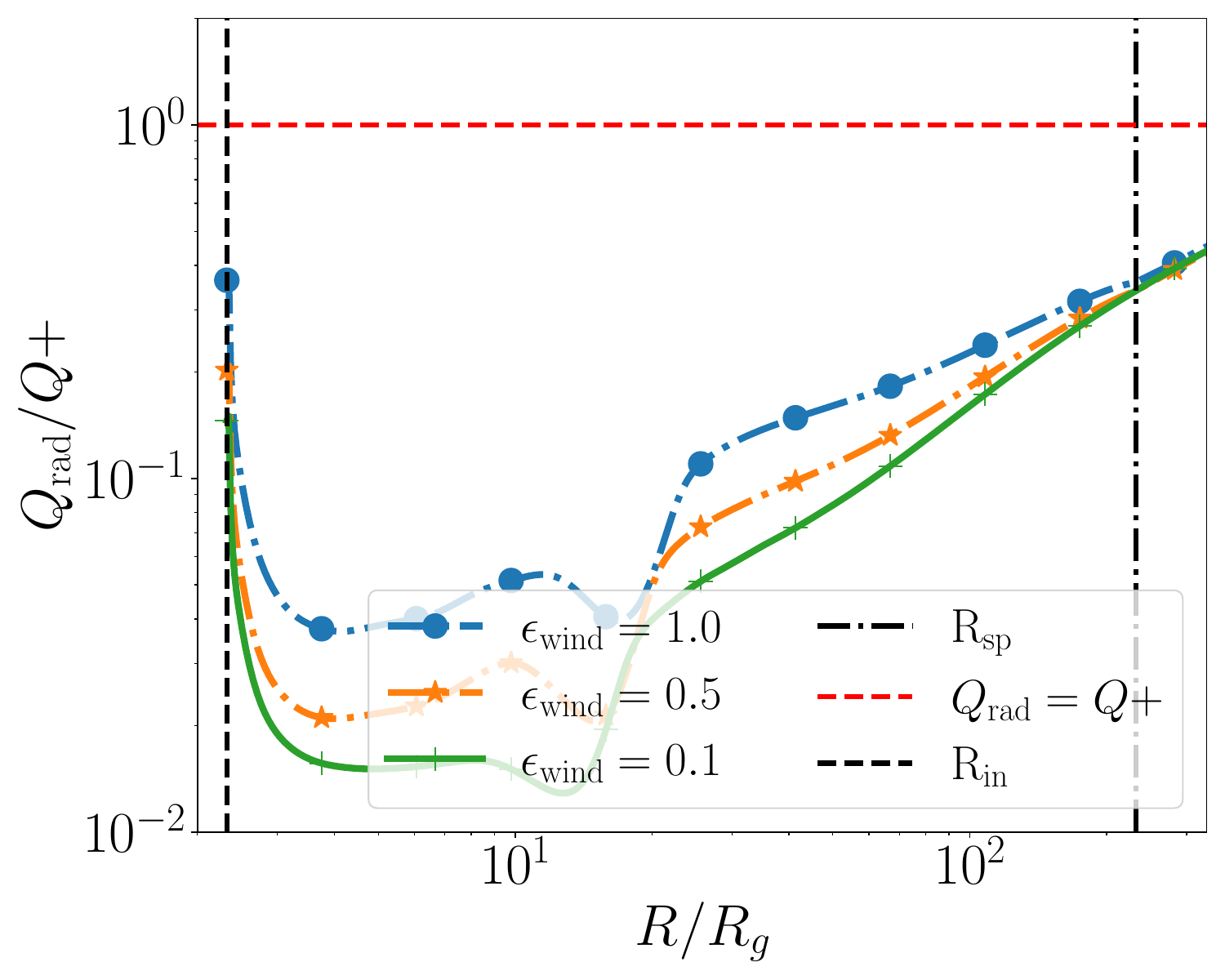}

    \caption{Plots of the ratio of advective cooling (left) and radiative cooling (right) to total heating as a function of radius for $\epsilon_{\rm wind} = 0.1, 0.5, 1.0$ with $\beta = 45^\circ$.}
    \label{fig:changing_ewind_2}
\end{figure*}

A reduced mass loss with reduced $\epsilon_{\rm wind}$ also produces a higher surface density, as shown in Figure \ref{fig:changing_ewind_1} (right panel). This difference is greatest within the warp region, where the surface density peak increases as $\epsilon_{\rm wind}$ decreases. At larger radii, the profiles begin to converge, indicating that changing the wind efficiency primarily modifies the inner disk. This higher surface density increases the optical depth of the disk and makes it more difficult for radiation to escape, increasing the importance of advective energy transport, which can be seen in Figure \ref{fig:changing_ewind_2}. As $\epsilon_{\rm wind}$ decreases, $Q_{\rm rad} / Q+$ becomes smaller and $Q_{\rm adv} / Q+$ approaches unity. %As is also clear, $Q_{\rm adv} / Q_{\rm rad}$ strongly increases, showing that, as the winds become less efficient in removing mass, advection becomes significantly more dominant.

\subsection{Misalignment angle versus $\dot{M}_{\rm in}$}
\label{sec:BetaVsMdotIn}

\begin{figure}
    \centering
    \includegraphics[width=1.0\linewidth]{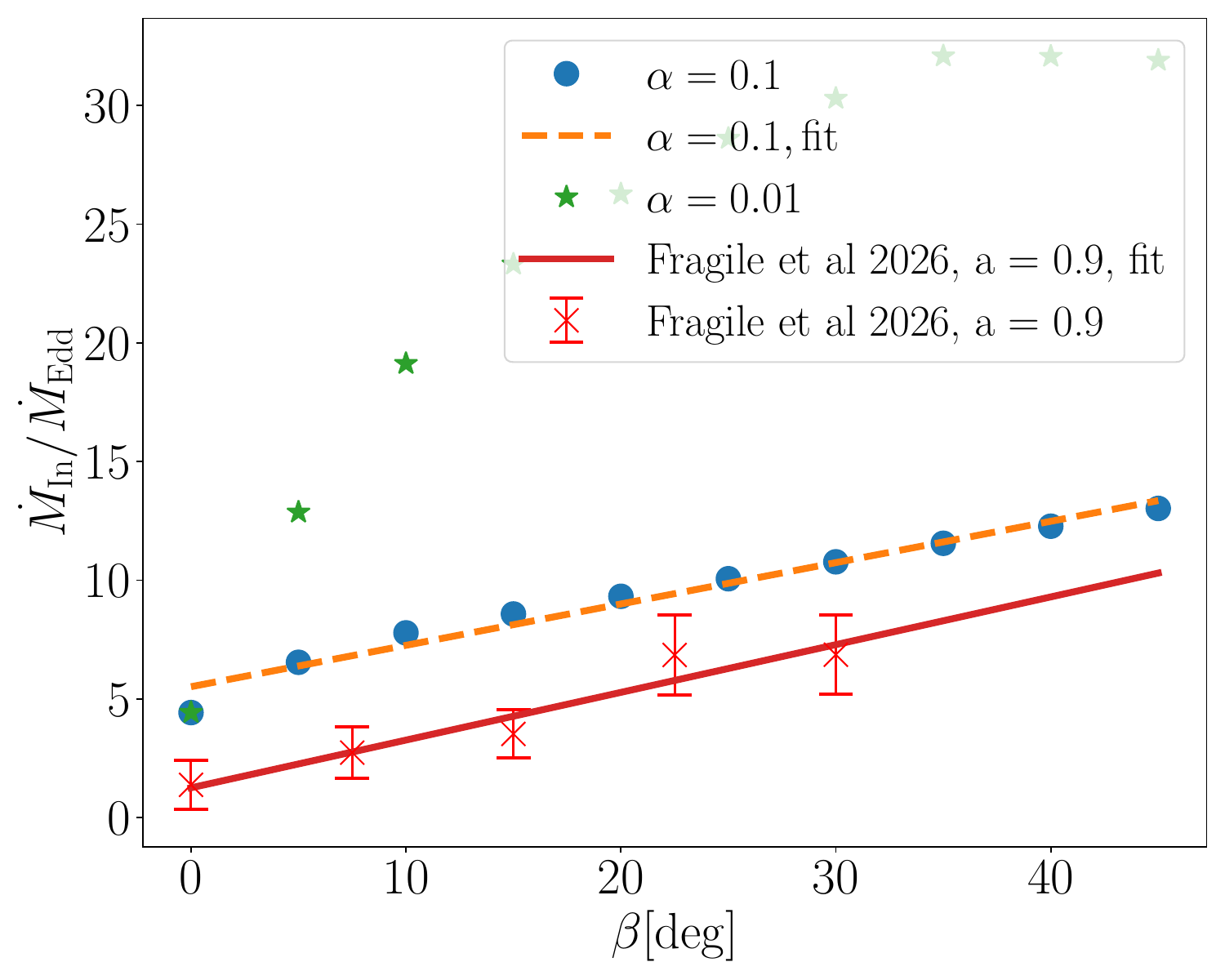}
    \caption{Mass accretion rate at the ISCO as a function of misalignment angle. Comparison with data taken from \citet{Fragile26} shows that our $\alpha = 0.1$ results are in closer agreement with the simulation data.}
    \label{fig:MdotIn}
\end{figure}

Figure \ref{fig:MdotIn} shows the mass accretion rate at the ISCO as a function of the misalignment angle $\beta$. As expected from the radial mass accretion rate profiles shown in Figure \ref{fig:MdotR} (left panel), increasing disk misalignment allows more mass to reach the innermost radii. The unwarped case ($\beta = 0^\circ$) shows the lowest value of $\dot{M}(R_{\rm in})$, and the value of $\dot{M}(R_{\rm in})$ increases approximately monotonically as the misalignment angle increases.

This strongly agrees with previous numerical work which showed that increasing tilt angle increases the amount of mass that reaches the innermost radii \citep[see Figure 5 in][]{Fragile26}. Figure \ref{fig:MdotIn} also shows that the response to misalignment depends strongly on $\alpha$. For $\alpha = 0.01$, the accretion rate at the ISCO rises steeply with tilt before plateauing at approximately $30 \dot{M}_{\rm Edd}$ above $\beta =35^\circ$, whereas $\alpha = 0.1$ only reaches $\sim 13 \dot{M}_{\rm Edd}$ by $\beta = 45^\circ$. This behavior is expected from Equation \ref{nu2onnu1}, where $\nu_2 / \nu_1 \propto \alpha^{-2}$. This means that decreasing $\alpha$ increases the relative vertical-shear viscosity and strengthens the warp-related redistribution of angular momentum. The model using $\alpha = 0.1$ reproduces the monotonic increase of $\dot{M}_{\rm in}$ with increasing misalignment and is closer to the GR-RMHD results than the model using $\alpha = 0.01$, suggesting that this result provides more accurate viscosity values within the assumptions of our model.

\section{Discussion and Conclusions}
\label{sec:discussion}

The results of our work show that introducing a warp can significantly alter the structure and energy transport within the inner regions of a super-Eddington accretion disk. In the unwarped limit, our model successfully reproduces the results from previous analytic super-Eddington disk solutions \citep[see][]{Lipunova_99, Poutanen_07}, including a sharp decline in the inward mass accretion rate due to wind losses, although we find that the disk slightly favors advection over radiation. 

Once a warp is introduced, our results begin to significantly diverge from the unwarped case. The warped models exhibit significantly higher $\dot{M}$ at the ISCO. Physically, this suggests that the warp-related torques and resulting angular momentum transport enhance the ability of the inner disk to trap radiation, reducing the local radiative efficiency and allowing more material to reach smaller radii. This is especially evident in the energy profiles. For non-zero warp angles, the ratio of $Q_{\rm adv} / Q_{\rm rad}$ exceeds unity in the inner disk, the ratio of $Q_{\rm adv} / Q+$ approaches unity, and $Q_{\rm rad} / Q+$ falls far below unity. Together, these results indicate that advection becomes the dominant cooling process within the inner disk. One consistent result is that all of these effects due to the warp are localized within the inner disk, specifically around the warp radius, while the various profiles tend to re-converge around the spherization radius, implying that the warp does not affect the outer disk.

These results are important in the broader context of rapid black hole growth. One of the main challenges in explaining the existence of SMBHs at high redshift is that radiative feedback in the form of winds might limit long-term accretion to values near  the Eddington rate \citep[see][]{Fragile_25}. In our model, as with the recent simulations by \citet{Fragile26}, we are able to reach super-Eddington rates of growth via a misalignment between the disc and black hole spin axes. Such misalignment will be a natural consequence of repeat feeding at a range of (potentially random) angles and naturally yields increased advection within the inner disk. 

One of the key advantages of our analytical approach is that it very rapidly allows us to explore the impact of a range of parameters which cannot be easily tested numerically. One of the major drawbacks however is that it is absent any complicated non-linear physics (e.g., MHD). Although we briefly compare our results to those in \citet{Fragile26}, future work will relax these assumptions through further direct comparisons with GR-RMHD simulations.

\section*{Acknowledgements}
A portion of this work was funded by a Royal Astronomical Society Summer Undergraduate Research Bursary and through NASA award No 80NSSC24K0900. I would like to thank the Royal Astronomical Society for their funding.

%%%%%%%%%%%%%%%%%%%%%%%%%%%%%%%%%%%%%%%%%%%%%%%%%%
\section*{Data Availability}

The code used for this work is available upon request to the lead author.



\bibliographystyle{mnras}
\bibliography{sample701} % if your bibtex file is called example.bib

%\begin{thebibliography}{99}
%\end{thebibliography}

%%%%%%%%%%%%%%%%%%%%%%%%%%%%%%%%%%%%%%%%%%%%%%%%%%

%%%%%%%%%%%%%%%%% APPENDICES %%%%%%%%%%%%%%%%%%%%%

\appendix

\section{Surface Density Function Derivation}
\label{appendA}
The following goes through the full derivation of equation (\ref{Sigma}) for surface density. Starting from equation \ref{eq:dotMR}, use the substitution
\begin{equation}
    y(R)=\nu_1 \Sigma(R) R^{1/2}
    \label{eq:y(R)}
\end{equation}
so that
\begin{equation}
    \frac{y}{\nu_1}=\Sigma(R)R^{1/2}.
\end{equation}
Then, using
\begin{equation}
    \dot{M}(R)
    =
    6\pi R^{1/2}
    \left[
        \frac{dy}{dR}
        +
        y
        \left(\frac{\nu_2}{\nu_1}\right)
        \frac{R}{3}
        \left|\frac{\partial \vec{l}}{\partial R}\right|^2
    \right] ~,
\end{equation}
we define
\begin{equation}
    h(R)=\left(\frac{\nu_2}{\nu_1}\right)\frac{R}{3}\left|\frac{\partial \vec{l}}{\partial R}\right|^2
\end{equation}
and
\begin{equation}
    g(R)=\frac{\dot{M}(R)}{6\pi R^{1/2}} ~.
    \label{eq:g(R)}
\end{equation}
The equation then becomes
\begin{equation}
    \frac{dy}{dR}+h(R)y=g(R) ~.
    \label{eq:dydR}
\end{equation}
This first-order linear ordinary differential equation is solved using the integrating factor:
\begin{equation}
    \mu(R)=\exp\left(\int^R_{R_{\rm in}} h(s)\,ds\right).
\end{equation}
Multiplying equation \ref{eq:dydR} through by $\mu(R)$ gives
\begin{equation}
    \mu(R)\frac{dy}{dR}+\mu(R)h(R)y=\mu(R)g(R),
\end{equation}
so that the left-hand side is
\begin{equation}
    \frac{d}{dR}\left[\mu(R)y(R)\right]
    =
    \mu(R)\frac{dy}{dR}+\mu(R)h(R)y.
\end{equation}
Therefore,
\begin{equation}
    \frac{d}{dR}\left[\mu(R)y(R)\right]=\mu(R)g(R).
\end{equation}
Integrating from the inner radius $R_{\rm in}$ to $R$,
\begin{equation}
    \mu(R)y(R)-\mu(R_{\rm in})y(R_{\rm in})
    =
    \int_{R_{\rm in}}^{R}\mu(R')g(R')\,dR'.
\end{equation}
Assuming a zero-torque inner boundary condition
\begin{equation}
    y(R_{\rm in})=0,
\end{equation}
this reduces to
\begin{equation}
    y(R)=\mu(R)^{-1}\int_{R_{\rm in}}^{R}\mu(R')g(R')\,dR'.
\end{equation}

Substituting back $y(R)=\nu_1\Sigma(R)R^{1/2}$ from equation (\ref{eq:y(R)}) gives
\begin{equation}
    \nu_1\Sigma(R)R^{1/2}
    =
    \mu(R)^{-1}\int_{R_{\rm in}}^{R}\mu(R')g(R')\,dR',
\end{equation}
and hence
\begin{equation}
    \Sigma(R)
    =
    \frac{\mu(R)^{-1}}{\nu_1 R^{1/2}}
    \int_{R_{\rm in}}^{R}\mu(R')g(R')\,dR'.
\end{equation}
Now substituting for $g(R')$ from equation (\ref{eq:g(R)}):
\begin{equation}
    \Sigma(R)
    =
    \frac{\mu(R)^{-1}}{6\pi \nu_1 R^{1/2}}
    \int_{R_{\rm in}}^{R}
    \frac{\dot{M}(R^\prime)}{R^{'{1/2}}}
    \mu(R')\,dR'.
\end{equation}

It is convenient to define the kernel:
\begin{equation}
    I(r,r') \equiv \frac{\mu(r')}{\mu(R)}
    =
    \exp\left(-\int_{r'}^{r} h(s)\,ds\right),
\end{equation}
so that the surface density may be written as:
\begin{equation}
    \Sigma(R)
    =
    \frac{1}{6\pi \nu_1 R^{1/2}}
    \int_{R_{\rm in}}^{R}
    \frac{\dot{M}(R^\prime)}{R'^{1/2}}
    I(R,R')\,dR'
\end{equation}

To make everything dimensionless, we multiply and divide by $\dot{M}_0$
\begin{equation}
    \Sigma(R) = \frac{\dot{M}_0}{6\pi \nu_1 R^{1/2}} \int_{R_{\rm in}}^{R} \frac{\dot{M} (R)}{\dot{M}_0} \frac{I(R,R^{\prime})}{\sqrt{R}} dR'
\end{equation}
and define a dimensionless function $f(r)$ as
\begin{equation}
    f(R) \equiv \frac{1}{2 \sqrt{R}} \int_{R_{\rm in}}^{R^{\prime}} \frac{\dot{M} (R)}{\dot{M}_0} \frac{I(R,R^{\prime})}{\sqrt{R}} dR^{\prime} ~,
\end{equation}
which when multiplied by $\dot{M}_0 / 3 \pi \nu_1$ returns the formula for $\Sigma(r)$.

\section{Infall Velocity Function Derivation}
\label{appendix:B}

Begin with Equation \ref{eq:mass_con} and multiply through by $R$ and move the wind term to the right side:
\begin{equation}
    \label{eq:appenB1}
    \frac{\partial}{\partial R}\left(R \Sigma V_R\right) = - {\frac{\partial \dot{m}_{\rm w}}{\partial R}}
\end{equation}

We then move onto Equation \ref{eq:angmom_con}:
\begin{equation}
\label{eq:appenB2}
    \begin{split}
     \frac{1}{R} \frac{\pa}{\pa R} \left(\Sigma V_R R^3 \Omega_K \vec{l}\right) + \frac{R^2 \Omega_K \vec{l}}{R} \frac{\partial \dot{m}_{\rm w}}{\partial R} =
    \\ \frac{1}{R} \frac{\partial}{\partial R}\left(\nu_1 \Sigma R^3  \vec{l} \ \frac{\partial \Omega_K}{\partial R}\right) + \frac{1}{R}\frac{\partial}{\partial R} \left(\frac{1}{2} \nu_2 \Sigma R^3 \Omega_K \frac{\partial \vec{l}} {\partial R}\right)
\end{split}
\end{equation}

Rewrite and expand the first term on the left side:
\begin{equation}
    \label{eq:appenB3}
    \begin{split}
    \frac{\pa}{\pa R}\left(\left(\Sigma V_R R \right)\left( R^2 \Omega_K \vec{l}\right) \right) = \\ \frac{R^2 \Omega_K \vec{l}}{R} \frac{\pa}{\pa R}\left(\Sigma V_R R \right) + \Sigma V_R  \vec{l} \frac{\pa \left(R^2 \Omega_K\right)}{\pa R} + \Sigma V_R R^2 \Omega_K \frac{\pa \vec{l}}{\pa R}
    \end{split}
\end{equation}
which you can then insert back into the left side of Equation \ref{eq:appenB2}:
\begin{equation}
    \label{eq:appenB4}
    \begin{split}
    \frac{R^2 \Omega_K \vec{l}}{R} \frac{\pa}{\pa R}\left(\Sigma V_R R \right) + \Sigma V_R \frac{\pa \left(R^2 \Omega_K\right)}{\pa R} \vec{l} \\ + \Sigma V_R R^2 \Omega_K \frac{\pa \vec{l}}{\pa R} + \frac{R^2 \Omega_K \vec{l}}{R} \frac{\partial \dot{m}_{\rm w}}{\partial R}
    \end{split}
\end{equation}
but from Equation \ref{eq:appenB1} we can cancel out the $\dot{m}_{\rm w}$ terms, so the angular momentum conservation equation then becomes:
\begin{equation}
    \label{eq:appenB5}
    \begin{split}
    \Sigma V_R \frac{\pa \left(R^2 \Omega_K\right)}{\pa R} \vec{l} + \Sigma V_R R^2 \Omega_K \frac{\pa \vec{l}}{\pa R} = \\ \frac{1}{R} \frac{\partial}{\partial R}\left(\nu_1 \Sigma R^3  \vec{l} \ \frac{\partial \Omega_K}{\partial R}\right) + \frac{1}{R}\frac{\partial}{\partial R} \left(\frac{1}{2} \nu_2 \Sigma R^3 \Omega_K \frac{\partial \vec{l}} {\partial R}\right)
    \end{split}
\end{equation}

Now we expand the right side of Equation \ref{eq:appenB5}. We first start with the first term on the right side:
\begin{equation}
    \label{eq:appenB6}
    \begin{split}
        \frac{1}{R} \frac{\partial}{\partial R}\left( \left(\nu_1 \Sigma R^3 \frac{\partial \Omega_K}{\partial R} \right) \vec{l}\right) = \\
        \frac{1}{R} \frac{\pa }{\pa R} \left(\nu_1 \Sigma R^3 \frac{\pa \Omega_K}{\pa R} \right) \vec{l} + \\
        \nu_1 \Sigma R^2 \frac{\pa \Omega_K}{\pa R} \frac{\pa \vec{l}}{\pa R}
    \end{split} 
\end{equation}
and then the second term on the right side:
\begin{equation}
    \label{eq:appenB7}
    \begin{split}
        \frac{1}{R}\frac{\partial}{\partial R} \left( \left(\frac{1}{2} \nu_2 \Sigma R^3 \Omega_K \right) \frac{\partial \vec{l}} {\partial R}\right) = \\
        \frac{1}{2R} \frac{\pa}{\pa R} \left(\nu_2 \Sigma R^3 \Omega_k \right) \frac{\pa \vec{l}}{\pa R} + \\
        \frac{1}{2} \nu_2 \Sigma R^2 \Omega_K \frac{\pa^2 \vec{l}}{\pa R^2}
    \end{split}
\end{equation}

Because $\vec{l}$ is a unit vector:
\begin{equation}
    \vec{l} \cdot \vec{l} = 1
\end{equation}
and
\begin{equation}
    \vec{l} \cdot \frac{\pa \vec{l}}{\pa R} = 0
\end{equation}
and
\begin{equation}
    \vec{l} \cdot \frac{\pa^2 \vec{l}}{\pa R} = - \left| \frac{\pa \vec{l}}{\pa R} \right|^2
\end{equation}

We can then take the dot product between $\vec{l}$ and Equation \ref{eq:appenB5} and using the above identities, we find that Equation \ref{eq:appenB5} becomes:
\begin{equation}
    \begin{split}
        \Sigma R V_R \frac{\pa (R^2 \Omega_K)}{\pa R} = \\
        \frac{\pa }{\pa R} \left(\nu_1 \Sigma R^3 \frac{\pa \Omega_K}{\pa R} \right) - \frac{1}{2} \nu_2 \Sigma R^3 \Omega_K \left| \frac{\pa \vec{l}}{\pa R} \right|^2
    \end{split}
\end{equation}

Dividing by $\Sigma R (\pa(R^2 \Omega_K) / \pa R)$ gives us the equation for infall velocity:
\begin{equation}
    V_R = \frac{\frac{\partial}{\partial R}\left(\nu_1\Sigma R^3 \frac{\partial \Omega_K}{\partial R}\right) - \frac{1}{2}\nu_2\Sigma R^3 \Omega_K\left|\frac{\partial \vec{l}}{\partial R}\right|^{2}}{\Sigma R \frac{\partial}{\partial R}\left(R^{2} \Omega_K\right)}
\end{equation}

% Don't change these lines
\bsp	% typesetting comment
\label{lastpage}
\end{document}